\documentclass[twocolumn]{aastex63}
\usepackage{mathtools}
\usepackage{bm}
\usepackage{enumerate}

\submitjournal{ApJ}

\usepackage{CJKutf8}

\newcommand{\abund}[2]{\ensuremath{[\mathrm{#1}/\mathrm{#2}]}}
\newcommand{\cfe}{\abund{C}{Fe}}

\newcommand{\metal}{\abund{Fe}{H}}

\newcommand{\rapo}{r_\ensuremath{\mathrm{apo}}}
\newcommand{\rperi}{r_\ensuremath{\mathrm{peri}}}

\def\vector#1{\mbox{\boldmath $#1$}}
\newcommand{\pa}{\partial}

\usepackage{amssymb} 
\usepackage{pifont} 
\shorttitle{The Atari Disk}
\shortauthors{Mardini et al.}

\begin{document}

\title{The Atari Disk, a Metal-Poor Stellar Population in the Disk System of the Milky Way}

\author[0000-0001-9178-3992]{Mohammad K.\ Mardini}
\affiliation{Kavli IPMU (WPI), UTIAS, The University of Tokyo, Kashiwa, Chiba 277-8583, Japan}
\affiliation{Institute for AI and Beyond, The University of Tokyo 7-3-1 Hongo, Bunkyo-ku, Tokyo 113-8655, Japan}

\author[0000-0002-2139-7145]{Anna Frebel}
\affiliation{Department of Physics and Kavli Institute for Astrophysics and Space Research, Massachusetts Institute of Technology, Cambridge, MA 02139, USA}

\author[0000-0002-7155-679X]{Anirudh Chiti}
\affil{Department of Astronomy $\&$ Astrophysics, University of Chicago, 5640 S Ellis Avenue, Chicago, IL 60637, USA}
\affil{Kavli Institute for Cosmological Physics, University of Chicago, Chicago, IL 60637, USA}

\author[0000-0003-3518-5183]{Yohai Meiron}
\affiliation{SciNet High Performance Computing Consortium, University of Toronto, 661 University Ave., Toronto, ON M5G\,1M1, Canada}

\author[0000-0002-8810-858X]{Kaley V. Brauer}
\affiliation{Department of Physics and Kavli Institute for Astrophysics and Space Research, Massachusetts Institute of Technology, Cambridge, MA 02139, USA}

\author[0000-0002-4669-9967]{Xiaowei Ou}
\affiliation{Department of Physics and Kavli Institute for Astrophysics and Space Research, Massachusetts Institute of Technology, Cambridge, MA 02139, USA}

\begin{abstract}
We have developed a chemo-dynamical approach to assign 36,010 metal-poor SkyMapper stars to various Galactic stellar populations. Using two independent techniques (velocity and action space behavior), $Gaia$ EDR3 astrometry, and photometric metallicities, we selected stars with the characteristics of the "metal-weak" thick disk population by minimizing contamination by the canonical thick disk or other Galactic structures. This sample comprises 7,127 stars, spans a metallicity range of $-3.50<${\metal}~$<-0.8$, and has a systematic rotational velocity of $\langle V_\phi\rangle=154$\,km\,s$^{-1}$ that lags that of the thick disk. Orbital eccentricities have intermediate values between typical thick disk and halo values. The scale length is $h_{R}=2.48^{+0.05}_{-0.05}$\,kpc and the scale height is $h_{Z}=1.68^{+0.19}_{-0.15}$\,kpc. 
The metallicity distribution function is well fit by an exponential with a slope of $\Delta\log{\rm N}/\Delta\metal=1.13\pm0.06$. 
Overall, we find a significant metal-poor component consisting of 261 SkyMapper stars with \metal$<-2.0$. While our sample contains only eleven stars with {\metal}~$\lesssim-3.0$, investigating the JINAbase compilation of metal-poor stars reveals another 18 such stars (five have {\metal}$<-4.0$) that kinematically belong to our sample. 
These distinct spatial, kinematic and chemical characteristics strongly suggest this metal-poor, phase-mixed kinematic sample to represent an independent disk component with an accretion origin in which a massive dwarf galaxy radially plunged into the early Galactic disk. Going forward, we propose to call the metal-weak thick disk population as the Atari disk, given its likely accretion origin, and in reference to it sharing space with the Galactic thin and thick disks. 

\end{abstract}

\correspondingauthor{Mohammad K.\ Mardini}
\email{m.mardini@ipmu.jp}


\section{Introduction} 
\label{sec:intro}

The existence of chemo-dynamically distinguishable components of the Galactic disk was first proposed several decades ago, in which the ``thick disk" was introduced as a distinct component of the Milky Way disk \citep[e.g.,][]{Gilmore.Reid1983}.
Many studies have investigated in detail the nature of this component, which is considered the ``canonical thick disk" by determining its age \citep[older than 8\,Gyr;][]{Kilic+2017}, velocity dispersion \citep[$\sigma_{z} \approx 35\,$km\,s$^{-1}$][]{Norris1993_thick_disk_velocity}, metallicity distribution (peaking at {\metal} $\approx -0.5$\footnote{\metal = $\log_{10}(N_{\text{Fe}}/N_{\text{H}})_{\star}-\log_{10}(N_{\text{Fe}}/N_{\text{H}})_{\sun}$}; \citealt{Kordopatis+2011}), and relative abundance ([X/Fe]) trends \citep[see;][]{Bensby2005}. In addition, a seemingly more metal-poor ({\metal} $<-0.8$) stellar population within this canonical thick disk was identified \citep{Norris1985_MWTD,Morrison1990_MWTD}, and termed the ``metal-weak thick disk" (MWTD) \citep[e.g.,][]{Chiba+2000}. 

While various properties of the canonical thick disk could be conclusively determined, the metal-weak thick disk remained insufficiently studied, likely due to its somewhat elusive nature. 
For example, several open questions remain regarding its nature -- what are the upper and lower {\metal} bounds characterizing the metal-weak thick disk? How did it form and evolve? Is it mainly the metal-poor tail of, and hence associated with, the canonical thick disk, or actually a separate component of the Milky Way's disk? 
Several clues from recent chemo-dynamical analyses \citet{Carollo2019,An2020} suggested the metal-weak thick disk as being independent from the the canonical thick disk, with plausibly distinct spatial, kinematic, chemical, and age distributions. 

Moreover, recent reports of very and extremely metal-poor stars being part of the Milky Way disk system has provided further insights into, and questions about, the formation of the Galactic disk system and the Milky Way itself \citep{Sestito2019,Carter2020, Cordoni+2020, Matteo2020, Venn2020}. The existence of these low-metallicity stars in the disk could be a signature of an early component of this disk system, assembled from a massive building block(s) entering the proto-Milky Way. 
Alternatively, these stars might have formed in the early disk system, which was later dynamically heated.

More generally, investigating thick disk origin scenarios through  metal-poor stellar samples of the disk may shed light on the nature and origin of the metal-weak thick disk; for instance, in gauging whether metal-weak thick disk stars have consistent behavior(s) or an implied origin that aligns with stars belonging to the canonical thick disk.
In this paper, we implement several approaches using kinematics derived from the \textit{Gaia} mission \citep{gaia+16, gaia+20} and photometric metallicities \citep{Chiti2020,Chiti2021} obtained from using public SkyMapper DR2 data \citep{owb+20} to select a clean and representative sample of metal-poor stars of the metal-weak thick disk. 

While addressing to what extent the metal-weak thick disk could be viewed as a component distinct from the canonical thick disk to learn about its early formation and evolution, we found that it is indeed characterizable as a distinct spatial, kinematic and chemical stellar component. While it appears independent of the thick disk, this disk component remains described by its low-metallicity stellar content, as originally envisioned with the description of ``metal-weak thick disk". To account for the different nature of this component, we propose to call it the Atari disk (with Atari \begin{CJK}{UTF8}{min} \textbf{辺り} \end{CJK} meaning ``in the neighborhood" or ``nearby" in Japanese), in reference to it sharing close space with the Galactic thin and thick disks. This paper explores a full characterization of the nature of the Atari disk which appears to have an accretion origin in which a massive dwarf galaxy plunged into the early Galactic disk.

\section{Sample Selection and Quality Assessment}

To build a representative sample of Atari disk/MWTD stars, we applied the following procedure. 
We used the photometric metallicity catalog presented in \citet{Chiti2021}, which provides metallicities ({\metal}) for $\sim 280,000$ relatively bright (g $\leqslant 17$) and cool (0.35 $<$ $g-i$ $<$ 1.20) giants using metallicity-sensitive photometry from SkyMapper DR2 \citep{owb+20}.
We then limited the sample to $g-i > 0.65$ and random metallicity uncertainties $< 0.5$\,dex, following \citet{Chiti2021_map}, to ensure a high-quality sample of photometric metallicities.
We cross-matched this sample with the early third data release of the \textit{Gaia} mission \citep[\textit{Gaia} EDR3,][]{gaia+20,Lindegren+2020a} to collect five-parameter astrometric solutions (sky positions: $\alpha$, $\delta$, proper motions: $\mu_{\alpha}\cos\delta$, $\mu_{\delta}$, and parallaxes: $\varpi$). 
For sources typical in our sample (e.g., brighter than G = 17\,mag), \textit{Gaia} EDR3 provides meaningfully more accurate astrometric measurements relative to \textit{Gaia} DR2. 
For instance, the parallax errors ($\Delta \varpi$) in our sample improve by $20\%$ and proper motion uncertainties improve by a factor of two.

In addition to these improvements, \citet{Lindegren+2020a} introduced several quality cuts for the selection of reliable astrometric solutions. 
We thus apply the following restrictions based on the \textit{Gaia} EDR3 quality flags which reduces our sample to 169,530 stars:

\begin{itemize}
    \item \texttt{astrometric\_excess\_noise} ($<$\,1\,$\mu$as):  Higher values might indicate that the astrometric solution for the target has failed and/or that the star is in a multiple-system for which the single object solution is not reliable\footnote{Without filtering on the astrometric excess noise, artefacts might present \citep[see Appendix C of][]{Lindegren2018}.}. This also accounts for the fact that our metallicity technique may fail for binaries.
    \item Parallax$\_$over$\_$error ($\geqslant$\,5): Ensures reliable distance measurements (i.e., 20\% uncertainty or better).
\end{itemize}

For reference, typical uncertainties in the parallaxes and proper motions of the resulting sample of stars are 0.01\,$\mu$as and 0.02\,$\mu$as\,yr$^{-1}$, respectively. 

To calculate the full space motions of our sample, line-of-sight radial velocities (RV) are required. 
About $\sim$7 million stars have RV measurements in the $Gaia$ DR2 catalog which is similar to what is available in \textit{Gaia} EDR3.
Yet only $\sim 19\,\%$ of our sample have any of these RV measurements. 
We apply an additional quality cut (\texttt{dr2\_radial\_velocity\_error} $ < 3.0$\,km\,s$^{-1}$) to conservatively select stars with reliable RV values. 
This results in a sample of 28,714 stars. To further increase the size of our sample, we collected additional high-quality RV data from other surveys. We acquired 311, 1581, 771, and 4905 unique measurements from the APOGEE DR16, LAMOST DR6, RAVE DR5, and GALAH DR3 surveys, respectively \citep{msf+17,lamost_3ref,Rave_5th, Buder+2021}. 
In case of stars having multiple spectroscopic RV measurements, we choose to keep the ones with the highest S/N. 
The final sample of our stars with available RV measurements increases to 36,010 stars after including these datasets. 

\begin{figure}[htbp!]
\includegraphics[width=0.5\textwidth]{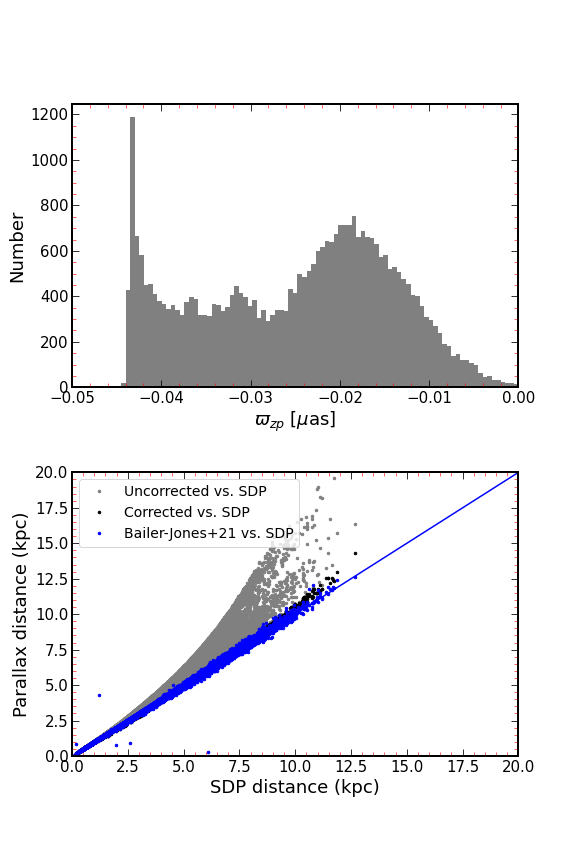}
\caption{Top panel: Distribution of our calculated parallax zero points for our final sample of 36,010 stars with available RV measurements. 
Bottom panel: Calculated parallax distances with the zero-point correction (black dots), without the zero-point correction (gray dots), and the \citet{Bailer-Jones2021} values (blue dots) as a function of the mean value distances calculated using a space density prior. 
The blue solid line represents the one-to-one relation.}
\label{fig:distances_Comparisons}
\end{figure}

We followed \citet{Lindegren+2020b} by assuming that additional parallax zero point ($\varpi_{zp}$) corrections are required for each star. 
These corrections utilize the magnitude, color, and ecliptic latitude of each source to compute an individual $\varpi_{zp}$ correction for each star in our sample.
For our sample, $\varpi_{zp}$ ranges from $-0.047$ to $0.004$\,$\mu$as, as shown in the upper panel of Figure~\ref{fig:distances_Comparisons}. 
We obtained corrected parallaxes ($\varpi_{corr}$) by subtracting the estimated $\varpi_{zp}$ from the $Gaia$ EDR3 parallaxes ($\varpi_{corr} = \varpi - \varpi_{zp}$).  

Stellar distances derived from directly inverting these corrected parallaxes ($d = 1 / \varpi$) should principally be reliable and not introduce additional biases \citep[e.g.,][]{Mardini_2019b,Mardini_2019a}. 
However, as an additional check, we calculated the distance of each star in our sample by implementing a Monte Carlo simulation with an exponentially decreasing space density prior as presented in \citet{Bailer-Jones}, which we label ``SDP" distances\footnote{At the time we started this project, the catalog in \citet{Bailer-Jones2021} was not public.}.
For this, we generated 10,000 realizations for each star assuming a normal distribution with $\varpi$ as the central value and the dispersion given by $\Delta \varpi$. 
We adopt the mean value as our final distance estimate. 
The lower panel of Figure~\ref{fig:distances_Comparisons} shows a direct comparison between distances calculated by inverting the parallax, our SDP approach, and the distances in \citet{Bailer-Jones2021}. 
For the parallax distances, we show two versions, one obtained without the zero-point correction, and one after the zero-point correction was applied. 

Out to 3\,kpc, all three distance measurements agree reasonably well. 
Beyond that, the un-corrected parallax distances are overestimated (the effect is prominent from 1-8\,kpc) compared to SDP distances, with the effect becoming worse at the largest distances.
However, the corrected parallax distances show excellent agreement with the SDP distances. 
We adopt the SDP distances for our entire sample, since they are more statistically vetted, though we note that the differences between those distances and the corrected parallax distances are minor.
Table~\ref{tab:Table_1} then lists $Gaia$ source ID, velocities, SDP distances, and orbital actions for each star of our final sample.

\begin{deluxetable*}{lrrrrrrrrrrr}
\caption{{Stellar Parameters and $Gaia$ Astrometric Solutions}}
\label{tab:Table_1}
\tabletypesize{\scriptsize}
\tablehead{
\colhead{source\_id} & 
\colhead{parallax}&
\colhead{$\mu_{\alpha}$ $cos(\delta)$} &
\colhead{$\mu_{\delta}$} &  
\colhead{RV}           & 
\colhead{l}        &
\colhead{b}       &
\colhead{d$_{SDP}$}& 
\colhead{{\metal}}      & 
\colhead{{L$_z$}}    &  
\colhead{{J$_r$}} &
\colhead{{J$_z$}}\\
\colhead{}&
\colhead{(mas)}	  &
\colhead{(mas\,yr$^{-1}$)}&
\colhead{(mas\,yr$^{-1}$)}&
\colhead{(km\,s$^{-1}$)}	&
\colhead{(deg)}     &
\colhead{(deg)}	&	
\colhead{(kpc)}   &  
\colhead{(dex)}	  &	
\colhead{(kpc km\,s$^{-1}$)} &
\colhead{(kpc km\,s$^{-1}$)} &
\colhead{(kpc km\,s$^{-1}$)}
}
\startdata
2334983060942097664&0.671&2.330&$-$28.582&70.27&37.245&$-$78.457&1.49&$-$0.86&542.99&389.25&192.81\\
4918304837496677248&0.304&4.711&$-$0.954&53.08&314.997&$-$56.947&3.31&$-$0.79&1305.19&26.08&163.45\\
4901413310941027072&1.125&32.894&$-$1.159&119.69&312.286&$-$52.664&0.89&$-$1.12&1015.84&158.48&198.71\\
2421111311440455168&0.198&$-$2.679&$-$9.637&$-$104.37&80.127&$-$71.531&5.09&$-$3.20&607.15&696.21&236.33\\
2422492847800684416&0.300&$-$0.500&$-$23.563&$-$110.44&83.613&$-$70.002&3.36&$-$2.42&$-$607.64&753.16&219.27\\
2339756040919188480&0.257&5.320&$-$7.972&69.20&47.902&$-$77.903&3.93&$-$1.25&605.90&238.43&356.76\\
4991401092065999360&0.376&0.444&3.777&20.18&328.375&$-$68.950&2.67&$-$0.81&2081.24&131.16&145.21\\
2340071806914929920&0.517&$-$4.974&$-$34.741&$-$97.15&50.676&$-$77.698&1.96&$-$1.97&$-$215.55&884.75&90.11\\
4995919432020493440&0.475&$-$2.472&$-$1.622&$-$34.00&333.978&$-$71.668&2.11&$-$0.89&1867.50&54.88&102.31\\
2341853840385868288&0.320&5.335&$-$19.603&$-$110.72&62.991&$-$76.206&3.15&$-$1.74&$-$508.89&376.26&137.58\\
2314830593353777280&0.727&30.731&$-$7.730&$-$18.87&13.964&$-$78.469&1.38&$-$0.92&852.38&467.33&36.53\\
4994799132751744128&0.354&0.902&$-$4.096&$-$4.87&331.310&$-$70.548&2.84&$-$0.88&1411.96&37.25&127.50\\
4688252950170144640&0.437&22.078&$-$2.191&$-$28.16&307.226&$-$41.562&2.29&$-$1.06&1303.86&427.46&60.41\\
2320839596198507648&0.604&5.558&$-$11.909&9.12&14.965&$-$78.575&1.66&$-$1.05&1118.50&145.25&49.47\\
2340104929702744192&0.185&$-$1.568&$-$12.358&$-$59.60&51.845&$-$77.765&5.44&$-$1.34&$-$49.71&827.14&429.74\\
4901401907804117632&0.693&43.595&$-$24.514&7.91&312.082&$-$52.667&1.44&$-$1.39&$-$19.81&874.88&94.39\\
2340079091179498496&0.732&$-$12.163&$-$11.067&27.42&50.690&$-$77.917&1.38&$-$0.93&1792.91&178.37&47.83
\enddata
\tablecomments{Parallax is the corrected parallax based on \citet{Lindegren+2020b}. $d_{\text{SDP}}$ is the mean value of the 10,000 realizations and {\metal} adopted from \citet{Chiti2021}. The complete version of Table~\ref{tab:Table_1} is available online only. A short version is shown here to illustrate its form and content. Distance and actions are rounded to two digits, but are given at full numerical precision in the online table.}
\end{deluxetable*}

\section{Derivation of Kinematic Parameters} \label{Sec:kinematic}

\subsection{Position and Velocity Transformation}

We transform  Galactic longitude ($l$), Galactic latitude ($b$), and distance to the Sun ($d$) to rectangular Galactocentric coordinates ($X,Y,Z$) using the following coordinate transformations:

\begin{align}\label{eq:rectangular_coordinate}
& X =  R_\odot - d\,\cos(l)\,\cos(b) \nonumber \\
& Y =  -d\,\sin(l)\,\cos(b) \\
& Z =  d\,\sin(b), \nonumber
\end{align}

\noindent where the Sun is located at R$_\odot =8.178 \pm 0.013$\,kpc from the Galactic center \citep{Gravity_Collaboration2019}; $X$ is taken to be oriented toward $l$=0$^{\circ}$, $Y$ toward $l$=90$^{\circ}$, and $Z$ toward the north Galactic pole.

We transform $\mu_{\alpha}\cos\delta$, $\mu_{\delta}$, and RV measurements to rectangular Galactic ($U,V,W$) velocities with respect to the Local Standard of Rest (LSR). $U$ is oriented toward the Galactic center, $V$ in the direction of Galactic rotation, and $W$ toward the North Galactic pole. We adopt the peculiar motion of the Sun ($U_{\odot} =11.1 \pm 0.72$\,km\,s$^{-1}$, $V_{\odot}= 12.24 \pm 0.47$\,km\,s$^{-1}$, and $W_{\odot}= 7.25 \pm 0.36$\,km\,s$^{-1}$) from \citet{Schonrich2010}, and a maximum height of $z_{\odot}= 20.8 \pm 0.3$\,pc of the Sun \citep{Bennett2019} above the plane. We take V$_{LSR}$ = $220$ km\,s$^{-1}$ from \citet{Kerr1986}\footnote{Using more recent values (e.g., 232.8 $\pm$ 3.0 km\,s$^{-1}$; \citealt{McMillan2017}) did not produce large discrepancies in the Galactic component classifications/membership. However, using such higher LSR value would shift the $<V_{\phi}>$ by ~10 km\,s$^{-1}$, which might create some confusion for the reader once we compare our calculated $<V_{\phi}>$ with literature values calculated using LSR = 220 km\,s$^{-1}$}.

We transform $U, V, W$ to velocities in cylindrical Galactocentric coordinates ($V_{R}, V_{\phi}, V_{z}$) using the following coordinate transformations:

\begin{align} \label{eq:cylindrical_velocities}
&V_{R}     = U \cos (\phi) + (V + V_{rot}) \sin (\phi) \nonumber\\
&V_{\phi}  = (V + V_{rot}) \cos (\phi) -  U \sin (\phi)  \\
&V_{z}     = W \nonumber 
\end{align}

\noindent Where $\cos$ ($\phi$) = $X/\sqrt{X{^2}+Y{^2}}$, $V_{rot}$ is the circular velocity of the LSR, $\sin$ ($\phi$) = $Y/\sqrt{X{^2}+Y{^2}}$, and objects with V$_{\phi} > 0$\,km\,s$^{-1}$ are retrograde.

\subsection{Orbital Parameters}

We used \texttt{galpy} and a scaled version of \texttt{MWPotential2014} potential \citep{Bovy2015} to derive orbital parameters (r$_{peri}$, r$_{apo}$, and Z$_{max}$) for each star. The modified \texttt{MWPotential2014} contains (i) a potential based on a virial mass of $M_{200} = 1.4 \times 10^{12}\,M_{\odot}$ instead of a canonical, shallower NFW profile, and (ii) a concentration parameter ($c = 8.25$) that matches the rotation curve of the Milky Way. This modification helps overcome an issue of erroneously identifying unbound stars, a known issue of the original \texttt{MWPotential2014} potential.

We define the total orbital energy as $E = (1/2) \vector{v}^2 + \Phi(\vector{x})$ and set $E = 0$ at a very large distance from the Galactic center. 
We define the eccentricity as $e = (\rapo - \rperi) / (\rapo + \rperi)$ and the vertical angular momentum component as $L_z = R \times V_{\phi}$ \citep[see][for more details]{galpy_Orbital_Parameters}. 
The distance from the Galactic center (cylindrical radius) is set by $R = \sqrt{X^{2}+Y^{2}}$.
We calculate these orbital parameters based on the starting point obtained from the observations via a Markov Chain Monte Carlo sampling method, assuming normally distributed input parameters around their observed values. 
We generate 10,000 realizations based on the observed input for each star to obtain medians and standard deviations of all kinematic parameters and to infer their values and associated uncertainties. We note that we use these orbital properties in all of the remaining Sections in the paper except in Section~\ref{sec:method1}, where we follow a separate approach to assign stars to their Galactic components.

\section{Identification of the Metal-Weak Thick Disk/Atari Disk and other Galactic Components}
\label{sec:selection}

The Galactic thick disk has been extensively studied as part of learning about the formation and evolution of the Milky Way system. Previous studies (beginning with \citealt{Gilmore.Reid1983} and the many others over the last several decades) all selected member stars assuming the thick disk to be a one-component, single population. However, it has long been suspected (e.g., \citealt{Norris1985_MWTD}) that a small portion of this canonical thick disk might actually be a separate, more metal-poor component that was eventually termed the ``metal-weak thick disk" \citep{Chiba+2000}. It was found that it has a mean rotational velocity $\langle V_{\phi}\rangle\,\sim 150$\,km\,s$^{-1}$ \citep[e.g.,][]{Carollo2019}. This presents a notable lag compared to the rotational velocity of the canonical thick disk with $\langle V_{\phi}\rangle \,\sim 180$\,km\,s$^{-1}$ \citep[e.g.,][]{Carollo2019}. Yet, more details remained outstanding to fully characterize this elusive body of metal-poor disk stars.

\citet{Beers+2014} suggested specific criteria to select MWTD stars, i.e. $Z_{max} \leqslant 3$\,kpc and $-1.8 \leqslant$ \metal $ \leqslant -0.8$. More recently \citet{Naidu2020} also suggested a MWTD selection criteria, namely $-2.5 \leqslant$ \metal $ \leqslant -0.8  \land (0.25<\rm{[\alpha/Fe]}<0.45) \land\ (J_{\phi}/\sqrt{J_{\rm{\phi}}^{2}+J_{\rm{z}}^{2}+J_{\rm{R}}^{2}} <-0.5)$. %
The chosen lower metallicity bounds aim to avoid possible contamination with the metal-poor halo. This prevents exploration of the potential existence of extremely low-metallicity stars typically associated with the halo within the MWTD (hereafter Atari). If the Atari disk has an accretion origin, it is principally expected that at least some extremely metal-poor stars should have been brought in from the early evolution of the progenitor.

\begin{figure}[htbp!]
\includegraphics[width=0.5\textwidth]{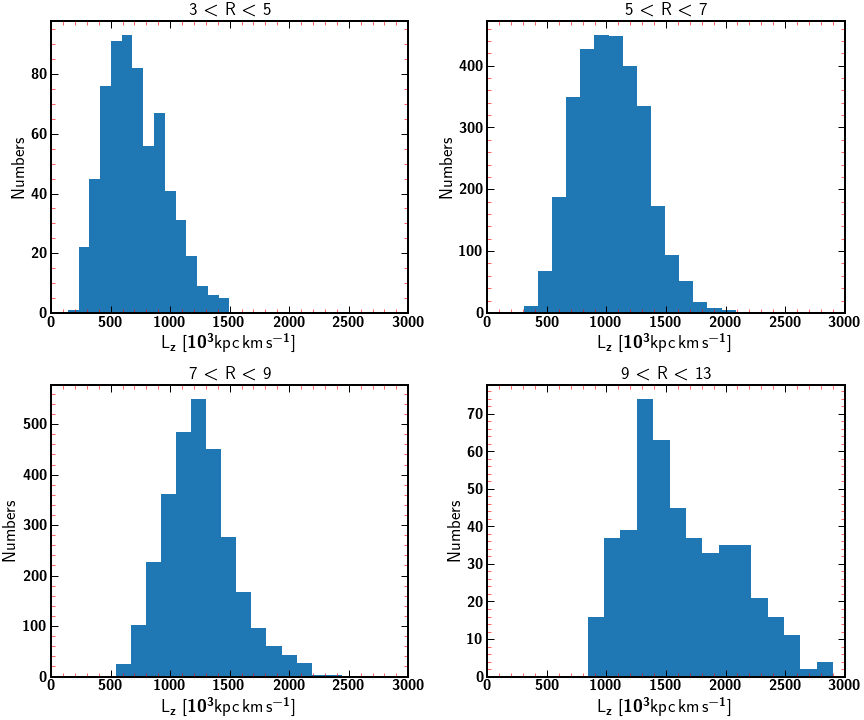}
\caption{Distribution of the angular momenta of our final SMSS Atari disk sample for different Galactic radii (R) cuts. The distributions have the same x-axis range to allow visual comparisons of the shift of the peak of the L$_z$ distributions between different R bins. The selection procedure for this sample is explained in Section~\ref{sec:method1} and~\ref{sec:method2}.
}
\label{fig:lz_distribution}
\end{figure}

Another selection has also been suggested by \citet{Carollo2019}, based on enhanced $\alpha$-abundances and an angular momentum of L${_z} \sim 1200$\,${\rm kpc}$\,km\,s$^{-1}$ (the less-prograde group in their Figure~1(a)) to characterize the Atari disk. However, using angular momentum as sole discriminator can only select stars within a given radial bracket as L${_z}$ varies as a function of Galactic radius $R$ (due to a roughly constant rotational velocity throughout the outer Galactic parts). For instance, for a sample restricted to the solar vicinity around R$\sim$8\,kpc, and using the suggested circular angular velocity of V$_{\phi} = 150$\,km\,s$^{-1}$, the resulting angular momentum is L${_z}$ = $1200$\,${\rm kpc}$\,km\,s$^{-1}$. But for a more distant sample, e.g. at $3<$R$<5$\,kpc, then L${_z}$ peaks at between $462$ and $770$\,${\rm kpc}$\,km\,s$^{-1}$ (see Figure~\ref{fig:lz_distribution}).

These different selection approaches show that it remains difficult to cleanly select Atari disk samples given that candidate stars have very similar chemical and kinematic properties to those of the canonical thick disk. In the following, we thus explore a different identification process to characterize the Atari disk, based on two different techniques (space velocities and behavior in action space) with the aim of selecting a representative, clean sample. We start by identifying stars in the thick disk using both of these methods, and then apply metallicity cut to isolate the Atari disk sample. This approach returns a sample of stars with kinematic properties in line with what was previously identified as the MWTD and thus allows us to more firmly establish the properties of this elusive Galactic component, including its low-metallicity member stars.

\subsection{Galactic Space Velocities Approach}
\label{sec:method1}
In order to select the traditional thin disk, thick disk and halo components, we adopt the kinematic analysis method presented in \citet{Bensby2003} that assumes the Galactic space velocities to have Gaussian distributions defined as follows:

\begin{align} \label{eq:velocities_distributions}
f(U,V,W)=k\cdot \textrm{exp}(&-\frac{(V_{\textrm{LSR}}-V_{\textrm{asym}})^2}{2\sigma_{V}^2}\nonumber\\
&-\frac{W_{\textrm{LSR}}^2}{2\sigma_{W}^2} -\frac{U_{\textrm{LSR}}^2}{2\sigma_{U}^2})
\end{align}

\noindent where
\begin{eqnarray}
k=\frac{1}{(2\pi)^{3/2}\sigma_{U}\sigma_{V}\sigma_{W}}\nonumber
\end{eqnarray}

\noindent The expressions $\sigma_{U}$, $\sigma_{V}$, and $\sigma_{W}$ denote the characteristic dispersions of each Galactic velocity component. The $V_{\textrm{asym}}$ denotes the asymmetric drift. We adopt these values from Table~1 in \citet{Bensby2003}.

To calculate the relative likelihood for a given star of being a member of a specific Galactic population, we take into account the observed number densities (thin disk $X_{D} = 0.8649$, thick disk $X_{TD} = 0.13$, and halo $X_{H} = 0.0051$) in the solar neighborhood vicinity (which we assume to be $\pm3$\,kpc from the Sun) as reported in \citet{Juric2008}. Therefore, the relative probabilities for the thick disk-to-thin disk (TD/D) and thick disk-to-halo (TD/H) ratios are defined as follows:

\begin{align} \label{eq:relative_probabilities}
&\textrm{TD/D} = \frac{X_{\textrm{TD}}\cdot f_{\textrm{TD}}}{X_{\textrm{D}}\cdot f_{\textrm{D}}} \nonumber\\
& \\
&\textrm{TD/H} = \frac{X_{\textrm{TD}}\cdot f_{\textrm{TD}}}{X_{\textrm{H}}\cdot f_{\textrm{H}}}  \nonumber
\end{align}

\begin{figure}[htbp!]
\includegraphics[width=0.5\textwidth]{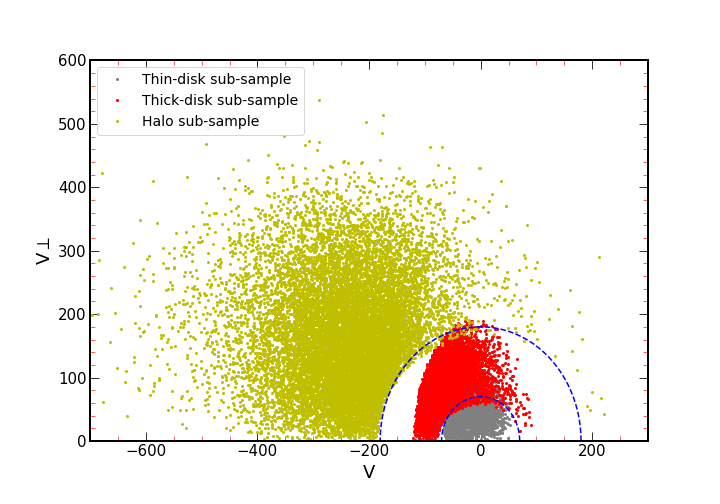}
\caption{Toomre diagram for our halo, thick disk, and thin disk stars in yellow, red, and gray points, respectively. Blue dashed curves denote $v_{{\rm tot}} = \sqrt{U_{{\rm LSR} } + V_{{\rm LSR}} + W_{{\rm LSR}} }= 70$ and $180\,{\rm km\,s^{-1}}$.}
\label{fig:probability_ratio}
\end{figure}

\noindent Following Eqs. \ref{eq:velocities_distributions} and \ref{eq:relative_probabilities}, we assign every star that has a membership probability of TD/D $> 2.0$ to the Galactic thick disk, while stars with TD/D $< 0.5$ are assigned to the Galactic thin disk. Furthermore, we exclude all stars with TD/H $< 10.0$ from the thick disk sample to minimize any possible contamination with halo stars. Our selection results in 10,588 thick disk stars, 2,571 thin disk stars, and 15,096 halo stars. 
Figure~\ref{fig:probability_ratio} shows a Toomre diagram of all these Galactic components, with typical halo stars having $v_{{\rm tot}} > 180\,{\rm km\,s^{-1}}$, and thick disk stars having $70\,{\rm km\,s^{-1}} < v_{{\rm tot}} < 180\,{\rm km\,s^{-1}}$. 
We note that discarding these low TD/H stars produces the small gap between the distributions of the thick disk (red) and halo (yellow) samples in Figure~\ref{fig:probability_ratio}.

\subsection{Orbital Properties Approach}\label{sec:method2}

In the previous subsection, we have identified thick disk stars by selecting its highly likely members (high relative probabilities) according to stellar velocities. 
Here, we develop another method to identify the traditional Galactic components based on the probability distribution functions of the action integrals of our sample, following the procedures presented in \citet{piffl+2014} and \citet{Posti+2018}.

The mass-normalized distribution function (DF) of the stellar halo
is assumed to have the following form:
\begin{align}
f_{\mathrm{halo}}(J_{r},J_{z},L_{z})=f_{0}\left[1+\frac{J_{r}+J_{z}+|L_{z}|}{J_{0}}\right]^{\beta_{*}}
\end{align}
where $f_{0}\approx0.09\,\mathrm{M}_{\odot}\mathrm{Gyr}^{3}\,\mathrm{kpc}^{-6}$
denotes a normalization constant that results in a total stellar halo mass\footnote{We note that adopting different total mass estimates \citep[e.g., $1.3\times 10^{9}$\,M$_{\odot}$;][]{Mackereth2020} would not change our halo membership assignment.} of $M_{\mathrm{halo}}=5\times10^{8}\,\mathrm{M}_{\odot}$.
The constant $J_{0}=511\,\mathrm{kpc}^{2}\,\mathrm{Gyr}^{-1}$ controls
the core in the center of the stellar halo, and the power law index $\beta_{*}=-4$
is chosen to set a reasonable density profile in the solar neighbourhood.

Most of the stellar mass is assumed to lie within the thin and thick disk components,
and follows quasi-isothermal DFs \citep{Binney2010} with the basic
form
\begin{equation}
\begin{split}
f_{\mathrm{disk}}(J_{r},J_{z},L_{z})=\frac{\Omega\Sigma\nu}{2\pi^{2}\sigma_{r}^{2}\sigma_{z}^{2}\kappa}\,[1+\tanh(L_z/L_0)]\,\times\\\exp(-\kappa J_{r}/\sigma_{r}^{2}-\nu J_{z}/\sigma_{z}^{2})\label{eq:quasi-isothermal-df}
\end{split}
\end{equation}
$\Omega$ is the circular frequency, $\kappa$ and $\nu$
are the radial and vertical epicycle frequencies, respectively, of
a planar circular orbit with angular momentum $L_{z}$ (these quantities are related to the potential through its spatial derivatives, see Chapter 3.2.3 of \citealt{Binney+2008}).
The surface density $\Sigma$ and the velocity dispersions ($\sigma_{r}$
and $\sigma_{z}$) are similarly functions of $L_{z}$, as they depend on the radius of a planar circular orbit in the potential.
We adopt L$_0 = 10\,\mathrm{kpc\,km\,s^{-1}}$ assuming that L$_0$ should not be bigger than the typical angular momentum of a star in the bulge/bar.
We then derive separate forms of DFs for each the thin and thick disk component through different adopted forms of the parameters (e.g., in $\sigma_r$, $\sigma_z$), exactly following \citet{piffl+2014}.

These DFs are calculated using the parameters given in \citet{Binney2016}\footnote{It is worth noting that adopting different literature parameters can meaningfully change the relative fraction of Galactic thin disk vs. thick disk stars}. 
These are the best fitting parameters given the assumed forms of the DFs, and other assumptions related to the kinematics of RAVE DR1 stars \citep{Steinmetz+2006} and the resulting mass distribution. 
The mass distribution has five components: thin and thick stellar disks, a gas disk, a flattened
stellar bulge, and a spherical dark matter halo (the stellar halo is neglected due to its relatively low mass), exactly following the form in \citet{piffl+2014}. We calculated the potential by numerically solving the Poisson equation given the mass distribution, and we then were able to evaluate the DFs for every J.
In the case of the thin disk, $\sigma_{r}$ and $\sigma_{z}$ are assumed to additionally depend on time (the velocity-dispersion functions increase with stellar age), and the DF is evaluated through a weighted time integration (again, following \citealt{piffl+2014}).

Note also that an additional order-unity multiplicative term in the
quasi-isothermal DF is found by \citet{Binney2010}. It is not used here
as that term is needed to control the asymmetry of probabilities with respect
to the direction of rotation (sign of $L_{z}$) that is not constrained
by \citet{piffl+2014}. Instead, \citet{Piffl2015} use a refined way to calculate the quasi-isothermal
DFs by iteratively inputting a newly calculated potential from the
DF back into Equation (\ref{eq:quasi-isothermal-df}) until convergence is achieved. 

In order to evaluate each of the three DFs for each star in our sample,
the actions have to be calculated. For internal consistency of this
method, we use the same potential that was used to derive the disk DFs.
Accordingly, we use a spherically symmetric ad-hoc approximation:
\begin{align}
\Phi_{\mathrm{approx}}(r)=-\Phi_{0,\mathrm{fit}}\frac{r_{\mathrm{fit}}}{r}\left[1-\frac{1}{\left(1+r/r_{\mathrm{fit}}\right)^{\beta_{\mathrm{fit}}-3}}\right]
\end{align}
This corresponds to the analytical potential of a $\beta$-model presented in \citet{Zhao1996}, where $\Phi_{0,\mathrm{fit}}=2.08\times10^{6}\,\mathrm{kpc}^{2}\:\mathrm{Gyr}^{-2}$,
$r_{\mathrm{fit}}=6.63\,\mathrm{kpc}$, and $\beta_\mathrm{fit}=3.148$. The approximate
potential is accurate to within 6\% everywhere inside the virial radius.
The actions are calculated using the formulae in \citet[chapter 3.5.2]{Binney+2008}

Finally, the probability to find a star in a phase space volume $\mathrm{d}^{3}\bm{J}$ around $\bm{J}_{i}$ is proportional to the value of the DF at this point divided by the total mass of the component. Therefore, relative probabilities are:
\begin{align}
\mathrm{TD/D}=\frac{f_{\mathrm{thick}}(\bm{J}_{i})/M_{\mathrm{thick}}}{f_{\mathrm{thin}}(\bm{J}_{i})/M_{\mathrm{thin}}}
\end{align}
and
\begin{align}
\mathrm{TD/H}=\frac{f_{\mathrm{\mathrm{thick}}}(\bm{J}_{i})/M_{\mathrm{thick}}}{f_{\mathrm{\mathrm{halo}}}(\bm{J}_{i})/M_{\mathrm{halo}}}
\end{align}
where $M_{\mathrm{thin}}=2.86\times 10^{10}\:\mathrm{M}_\odot$, $M_{\mathrm{thick}}=1.17\times 10^{10}\:\mathrm{M}_\odot$, and $M_{\mathrm{halo}}=5\times10^{8}\,\mathrm{M}_{\odot}$.

Using this approach and the same probability thresholds as in the previous section results in the selection of 15,521 thick disk stars, 3,278 thin disk stars, and 15,289 halo stars. These results are in good agreement; for example, the two methods select the main bulk (more than $\sim 87\%$) of each Galactic component obtained by the other selection technique in Section~\ref{sec:method1}. 
To construct a clean Atari disk sample, we then adopt an inclusion method by first selecting all thick disk stars that are common to both selection methods. Then, we only include stars with photometric \metal $< -0.8$, following the upper limit of the metallicity criteria in \cite{Beers+2014} and \cite{Naidu2020} to isolate the Atari disk. This results in a sample of 7,127 stars, which we hereby refer to as the Atari sample. We find that 261 stars in our Atari disk sample have \metal~$\leq -2.0$.

We decided to further assess the quality of our Atari disk sample via an independent check of our selection procedure, based on the spatial distribution of our Atari disk sample. 
We first considered the Z$_{max}$ distribution of the sample to identify any outliers (stars with high Z$_{max}$) that can plausibly be associated with the halo. 
The halo becomes more pronounced at Z$_{max} > 3$\,kpc, while the thin disk is confined to Z$_{max} < 0.8$\,kpc. 
The vast majority of our Atari disk sample lies in the range of $0.8 \leq$ Z$_{max} \leq 3$\,kpc which suggests that this sample predominantly includes objects not belonging to the halo or thin disk but rather in between, and thus more consistent with the thick disk.

As a second check, we then computed orbital histories for the past 10\,Gyr for all stars following \citet{Mardini_2020} to further validate their stellar membership to the Atari disk.
Again, the vast majority of stars have orbital properties similar to that of the canonical thick disk. This agrees with what was suggested by the Z$_{max}$ values, that contamination from the metal-poor halo or thin disk is low.

We do find that 439 stars in our sample of 7,127 Atari disk stars lie outside of the 0.8\,kpc $\leq$ $Z_{max} \leq 3.0$\,kpc range.
We find that the long-term orbits of these stars largely reflect thick disk characteristics, i.e., their orbital energies of E $<-0.9 $\,km$^2$\,s$^{-2}$, eccentricity of $0.3<$ e $<$ 0.5, and distances of $\rapo < 5$\,kpc very much align with thick disk kinematics. 
Only 137 stars of these 439 outlier stars have orbital histories not consistent with the thick disk. 
We do, however, find that all these stars have putative thick disk membership via evidenced by their TD/D ratios, as derived in Sections~\ref{sec:method1} and~\ref{sec:method2}, ranging from $10$ to $10^5$ and from $10$ to $10^8$, respectively. 

We thus conclude that our Atari disk sample is clean at the 98\% level, given the 137/7127 stars that do not have orbital histories consistent with thick disk-like motions.
Accordingly, at most relatively few stars with halo-like or thin disk-like kinematics are coincidentally selected by our technique. 
Considering this a representative sample of the Atari disk, we now assess various characteristics to describe this elusive Galactic component.

\subsection{Simple Atari disk star selection recipe}\label{simple}

The bulk of our Atari disk sample has unique kinematic parameters unassociated with the general properties of the canonical thick disk. For example, the canonical thick disk is known to have a rotational velocity $V_{\phi} \approx 180$\,km\,s$^{-1}$, which lags the V$_{LSR}$ by $\sim40$\,km\,s$^{-1}$. But our Atari disk sample has rotational velocity $V_{\phi} = 154$\,km\,s$^{-1}$. Also, $\sim 20\%$ of our Atari disk sample have orbital eccentricities above the typical range of orbital eccentricities reported in the literature for the canonical thick disk (see Table~\ref{tab:populations}).

Given the complex and involved nature of our selection procedures in Sections~\ref{sec:method1} and~\ref{sec:method2}, we also attempted to develop a simplified procedure that would allow the selection of Atari disk stars from other existing and future stellar samples with more ease. We suggest the following. Stars that fulfil the criteria

\metal $\ < -0.8 \ \land \ Z_{max} < 3.0 \ \land \ (J_{\phi}/\sqrt{J_{\rm{\phi}}^{2}+J_{\rm{z}}^{2}+J_{\rm{R}}^{2}} <-0.98) \ \land \ ( 0.3 < e < 0.6) \ \land \ (140 < V_{\phi} < 160)$

will be Atari disk stars with high likelihood, albeit not yield an all-encompassing sample of Atari disk stars. 
Applying these criteria to our initial SMSS sample (36,010 stars), we find that 84\% of the stars recovered from our simple selection are also in the Atari disk sample. We investigated the nature of the remaining 16\% of stars found with the simple selection recipe. Our calculated membership probabilities suggest equal probability of thin and halo of these contaminants.

\section{Properties of the Atari disk}\label{sec:Atari_properties}

In this Section, we aim to establish the kinematic properties of the Atari disk using our representative sample as selected in the previous section. 
Specifically, we investigate the scale length, scale height, and correlations between several variables (e.g., metallicity, eccentricity, rotational velocity) to characterize the nature of this component.
Table~\ref{tab:populations} lists our derived properties of the Atari disk, along with those of other galactic populations for comparison. 

\begin{table*}[ht!]
	\begin{center}
		\centering
		\caption{Orbital properties of the Galactic thin disk, thick disk, and inner halo}
		\label{tab:populations}
		\begin{tabular}{lccccr}
			\hline
			\hline	
			 Parameter & unit & Thin disk &  Thick disk & Inner halo & Atari disk\\
			\hline	
		     $h_{R}$&   (kpc)  &2.6\tablenotemark{a} - 3.0\tablenotemark{b}  & 2.0\tablenotemark{b} - 3.0\tablenotemark{c} & \nodata &  2.48 $\pm$ 0.05 \\
			 $h_{Z}$&   (kpc) &0.14\tablenotemark{d} - 0.36\tablenotemark{e}  & 0.5\tablenotemark{e} - 1.1\tablenotemark{e} & \nodata & 1.68$^{+0.19}_{-0.15}$ \\
			 $<V_{\phi}>$&   (km s$^{-1}$)               &  208\tablenotemark{e}  & 182\tablenotemark{g}&   0\tablenotemark{f}&  154 $\pm$ 1\\
		     Z$_{max}$&   (kpc)                         & $< 0.8$\tablenotemark{h}  & $0.8$ - $3.0$\tablenotemark{g} & $> 3.0$\tablenotemark{g} &  $<$ 3.0\\
		     e&  \nodata                                       & $< 0.14$\tablenotemark{g}  & 0.3 - 0.5\tablenotemark{g} & $>$ 0.7\tablenotemark{g}& 0.30 - 0.7\\
			\hline
		\end{tabular}
	\end{center}
	\tablenotetext{}{References are as follows: (a): \citealt{Juric2008}; (b): \citealt{Li2017}; (c): \citealt{Li2018}; (d): \citealt{Sanders2015}; (e): \citealt{Recio2014}; (f): \citealt{Carollo2010}; (g): \citealt{Lee2011}; (h):  \citealt{Anders2014}. }
\end{table*}

\subsection{Scale Length}
\label{sec:scalelength}

Measurements of the scale length ($h_{R}$) and scale height ($h_{Z}$) are important to trace the structure, size, mass distribution, and radial luminosity profile of the Galactic disk components \citep[e.g.,][]{Dehnen1998}. 
In order to calculate $h_{R}$ and $h_{Z}$ of our Atari disk sample, we solve the fundamental collisionless Boltzmann equation of axisymmetric systems, which is expressed as the following \citep[see equation 4.12;][]{Binney+2008}:

\begin{align} \label{eq:boltzmann}
\frac{\pa f}{\pa t}  	& + v_R \frac{\pa f}{\pa R} + \frac{v_\phi}{R^{2}} \frac{\pa f}{\pa \phi} + v_z \frac{\pa f}{\pa z} - \left(\frac{\pa \Phi}{\pa R} - \frac{v_\phi^2}{R^{3} } \right) \frac{\pa f}{\pa v_R} \nonumber \\
			  	& - \frac{\pa \Phi}{\pa\phi} \frac{\pa f}{\pa v_\phi} - \frac{\pa \Phi}{\pa z}\frac{\pa f}{\pa v_z} = 0,
\end{align}

\noindent where $f$ is the number of objects in a small volume, and $\Phi$ is the gravitational potential. 
It is then convenient to derive the Jeans equation from the Boltzmann equation in the radial and Z-component directions as the following \citep[see equation 9;][]{Gilmore1989}:

\begin{align} \label{eq:Jeans1}
\rho K_{R}= \frac{1}{R} \frac{\pa (R\rho\sigma^{2}_{V_{R}})}{\pa R} + \frac{\pa (\rho\sigma^{2}_{V_{R,Z}})}{\pa Z} - \frac{\rho \sigma^{2}_{V_{\phi}}}{R}- \frac{\rho}{R} \bar{V_{\phi}}^{2}
\end{align}

\begin{align} \label{eq:Jeans2}
\rho K_{Z}= \frac{\pa (\rho\sigma^{2}_{V_{Z}})}{\pa Z} + \frac{1}{R} \frac{\pa (R\rho\sigma^{2}_{V_{R,Z}})}{\pa R}
\end{align}

\noindent where $\rho(R,Z)$ is the space density of the stars in the thick disk, and $K_{R}$= $\frac{\pa \phi}{\pa R}$, and $K_{Z}$= $\frac{\pa \phi}{\pa Z}$ are the derivatives of the potential. Assuming an exponential density profile, the radial Jeans equation can be rewritten as  follows \citep{Li2018}:

\begin{align} \label{eq:Jeans3}
\frac{\sigma^{2}_{V_{\phi}}}{\sigma^{2}_{V_{R}}}-2+\frac{2R}{h_{R}}-\frac{V_{c}^{2} -\bar{V_{\phi}}^{2} }{\sigma^{2}_{V_{R}}}+\frac{\sigma^{2}_{V_{Z}}}{\sigma^{2}_{V_{R}}} = 0
\end{align}

\noindent where $h_{R}$ is the scale length. 
By substituting our calculated velocity dispersions from the Atari disk sample, within $\approx$ 3\,kpc of the Sun in the cynlindrical $R$ coordinate and $\approx$ 2\,kpc above or below the Galactic plane, (6,347 stars) into Equation~\ref{eq:Jeans3}, we obtain a radial scale length of $h_{R} =2.48$\,kpc. Calculating the scale length using different metallicity bins shows a small increase from 2.38\,kpc among the higher metallicity stars up to 2.91\,kpc for the low-metallicity stars. The results are detailed in Table~\ref{tab:lengths_and_heights}. 
In general, these results point to the Atari disk being comparable in size in the radial direction to the thick and thin disk. 
For reference, the scale length of the canonical thick disk has been measured as 2.0\,kpc \citep{Bensby2011}, 2.2\,kpc \citep{Carollo2010}, and 2.31\,kpc \citep{Sanders2015}, although larger values have also been reported previously \citep{Chiba+2000,Jong2010}.  
Thin disk values refer to an overall similar spatial distribution although it is likely somewhat more extended ($h_{R} > 3.0$\,kpc; e.g., \citealt{Bensby2011,Sanders2015}). 
See Table~\ref{tab:populations} for further details.

\subsection{Scale Height}
Assuming an exponential density distribution and constant surface density ($\sigma^{2}_{V_{R,Z}} \approx 0$)\footnote{$\sigma^{2}_{V_{R,Z}}$ is negligible small compared to the remaining term in Eq.(14)}, as described in \citet{Gilmore1989}, Equation~\ref{eq:Jeans2} can be rewritten as follows:

\begin{align} \label{eq:Jeans4}
\frac{\pa \ln{\sigma^{2}_{V_{Z}}}}{\pa Z} - \frac{1}{h_{Z}}  + \frac{K_{Z}}{\sigma^{2}_{V_{Z}}}=0
\end{align}

\noindent where $h_{Z}$ is the scale height. 
By substituting $K_{Z} = 2\pi G \times 71 M_{\odot}$\,pc$^{-2}$ at $|$z$|$ = 1.1\,kpc (see equation 4 in \citealt{Kuijken1991}), relevant velocity dispersions, and gradients into Equation~\ref{eq:Jeans4}, the scale height can be obtained.

We applied this technique to derive scale heights for both the original velocity-selected sample of 7,451 stars (see Section~\ref{sec:method1}) as well as the action-selected sample of 10,351 stars, using the same spatial selection as in Section~\ref{sec:scalelength}. By design, the velocity selection method employed in our study sets out to select stars roughly within the spatial distribution of the thick disk (in the z-direction by using $\sigma_{\text{W}}=35$\,km\,s$^{-1}$; see Section~\ref{sec:method1}). 
This might lead to a bias when attempting to use the velocity-selected sample to determine the scale height. Table~\ref{tab:lengths_and_heights} shows the results. Using the action-selection sample, we then derive 1.68\,kpc for the scale height of the Atari disk. Restricting the sample to stars with $-1.2<\metal<-0.8$, we find $h_{Z} = $1.92\,kpc. 
However, stars with $-1.5<\metal<-1.2$ suggest a lower value of $\sim1.37$\,kpc. 
Stars with even lower metallicity once again follow a wider distributions with larger scale heights. However, the larger uncertainty associated with the calculated $h_{Z}$ in this metallicity bin comes from the low number of stars to calculate the slope of the velocity dispersion (see term 1 of Equation~\ref{eq:Jeans4}) accurately. We also investigated the idea that these different $h_{Z}$ values might be due to possible contamination from other accreted substructures. To address this question, we investigated the E-L$_z$ space of each of the used different {\metal} bins. Overplotting these E-L$_z$ distributions on our Figure~\ref{fig:E_Lz_space} suggests no significant overlap with any of the other accreted substructures. Also, we performed a 2D Gaussian mixture model fitting in E-L$_z$ space for each of these {\metal} bins and found that the E-L$_z$ distribution in each {\metal} bin could be reasonably fit by one Gaussian. This suggests no obvious substructure contaminating our sample at various {\metal} bins.

At face value, the $h_{Z}$ values calculated for the action-selected sample are about 0.2 to 0.5\,kpc larger than what we find for the velocity-selected sample, as can be seen in Table~\ref{tab:lengths_and_heights}. While there is a small change in $h_{Z}$ for the highest metallicity bin ($-1.2<\metal<-0.8$) compared to the whole Atari disk sample, the low number of stars in the other metallicity bins (i.e., large uncertainties) refrain us from determining any scale height gradient with metallicity.
For comparison, using a chemo-dynamical selection, \citep{Carollo2010} find a scale height of $h_{Z} = 1.36$\,kpc. Their sample had \metal$<-1.0$ and ranging down to below \metal$=-1.5$. Such low value corresponds to what we obtain for our metallicity bin of $-1.5<\metal<-1.2$. Based on our comprehensive analysis, this value may well depend on the metallicity of the chosen sample. 

To further quantify the bias introduced by the velocity method, we reran our analysis with an increased $\sigma{_z}=45$\,km\,s$^{-1}$ and $\sigma{_z}=55$\,km\,s$^{-1}$. Our intention was to learn whether increasing the initial spatial distribution would impact the scale height to the extent of matching that of the action method. While choosing $\sigma{_z}=55$\,km\,s$^{-1}$ did indeed result in scale height increases, the values of the action-selected sample were not entirely reached. At the same time, however, the halo contamination rate drastically increased, suggesting that loosening the velocity selection criterion was detrimental to our overall science goal of accurately selecting a high-confidence sample of Atari disk stars. To avoid this bias, we thus chose to use the scale height value obtained from the action-selection sample only. For the remainder of the analysis, we then kept the common sample as originally selected with $\sigma{_z}=35$\,km\,s$^{-1}$. 

Interestingly, a scale height of $h_{Z} \sim1.7$\,kpc derived for the whole metallicity range is significantly more extended than what is measured for the canonical thick disk having $h_{Z} \sim0.5$ to 1\,kpc. More recent papers have reported progressively shorter scale heights (see Table~\ref{tab:populations}) which suggests that the scale height of the Atari disk is could be up to three times that of the thick disk. Considering a more matching metallicity range of these two populations, the Atari disk scale height for stars with $-1.2<\metal<-0.8$ is $\sim$1.9\,kpc which is about two to four times that of the thick disk. But even the lowest value derived for stars with $-1.5<\metal<-1.2$ of $\sim1.4$\,kpc is still more than the scale height of the thick disk.
Values for other metallicity bins can also be found in Table~\ref{tab:lengths_and_heights}, for additional comparisons. 
Overall, this robustly suggests the Atari disk to be generally significantly more extended than the thick disk in the the z-direction.

\begin{deluxetable}{lcccc}\label{tab:lengths_and_heights}
\tabletypesize{\scriptsize}
\tablecaption{Scale Lengths and Scale Heights for Different Metallicity Bins}
\tablehead{\colhead{Sample} &\colhead{Metallicity bin} & \colhead{$N_{\rm stars}$} &
  \colhead{scale length} & \colhead{scale height} \\
  & & & (kpc) & (kpc)}
\startdata
Common     &$-1.2\le {\rm [Fe/H]}<-0.8$ & 4,868 & \boldmath{$2.38^{+0.05}_{-0.05}$} & $1.75^{+0.17}_{-0.14}$\\
           &$-1.5\le {\rm [Fe/H]}<-1.2$ & 835   & \boldmath{$2.62^{+0.09}_{-0.09}$} & $1.36^{+0.31}_{-0.23}$\\
           &$-1.8\le {\rm [Fe/H]}<-1.5$ & 314   & \boldmath{$2.98^{+0.15}_{-0.15}$} & $1.41^{+1.37}_{-0.52}$\\
           &$-3.5\le {\rm [Fe/H]}<-1.8$ & 268   & \boldmath{$2.91^{+0.16}_{-0.16}$} & $2.03^{+2.82}_{-0.95}$\\
           &$-3.5\le {\rm [Fe/H]}<-0.8$ & 6,347 & \boldmath{$2.48^{+0.05}_{-0.05}$} & $1.67^{+0.20}_{-0.16}$\\
\hline
Velocity   &$-1.2\le {\rm [Fe/H]}<-0.8$ & 5,570 & $2.64^{+0.04}_{-0.04}$ & $1.43^{+0.18}_{-0.15}$\\
           &$-1.5\le {\rm [Fe/H]}<-1.2$ & 993   & $3.30^{+0.10}_{-0.10}$ & $1.15^{+0.19}_{-0.16}$\\
           &$-1.8\le {\rm [Fe/H]}<-1.5$ & 414   & $4.00^{+0.11}_{-0.12}$ & $1.33^{+1.14}_{-0.48}$\\
           &$-3.5\le {\rm [Fe/H]}<-1.8$ & 394   & $4.14^{+0.09}_{-0.09}$ & $1.68^{+1.32}_{-0.56}$\\
           &$-3.5\le {\rm [Fe/H]}<-0.8$ & 7,451 & $3.00^{+0.05}_{-0.05}$ & $1.39^{+0.19}_{-0.16}$\\
\hline
Action     &$-1.2\le {\rm [Fe/H]}<-0.8$ & 7,694 & $2.51^{+0.04}_{-0.04}$  & \boldmath{$1.92^{+0.17}_{-0.15}$}\\
           &$-1.5\le {\rm [Fe/H]}<-1.2$ & 1,497 & $2.95^{+0.08}_{-0.08}$  & \boldmath{$1.37^{+0.32}_{-0.22}$}\\
           &$-1.8\le {\rm [Fe/H]}<-1.5$ & 707   & $3.30^{+0.12}_{-0.12}$  & \boldmath{$1.63^{+1.19}_{-0.49}$}\\
           &$-3.5\le {\rm [Fe/H]}<-1.8$ & 639   & $3.38^{+0.12}_{-0.12}$  & \boldmath{$2.08^{+3.15}_{-1.10}$}\\
           &$-3.5\le {\rm [Fe/H]}<-0.8$ &10,351 & $2.70^{+0.04}_{-0.04}$  & \boldmath{$1.68^{+0.19}_{-0.15}$}\\
\hline
\hline
\multicolumn{4}{l}{Additional metallicity bins}\\
\hline
\hline
Common     &$-0.9\le {\rm [Fe/H]}<-0.8$ & 1,960 & $2.32^{+0.06}_{-0.06}$ & $1.64^{+0.47}_{-0.30}$\\
           &$-1.1\le {\rm [Fe/H]}<-0.9$ & 2,161 & $2.41^{+0.06}_{-0.06}$ & $2.03^{+0.87}_{-0.48}$\\ 
           &$-1.3\le {\rm [Fe/H]}<-0.8$ & 5,264 & $2.40^{+0.04}_{-0.05}$ & $1.73^{+0.18}_{-0.15}$\\ 
           &$-3.5\le {\rm [Fe/H]}<-1.1$ & 1,964 & $2.73^{+0.07}_{-0.07}$ & $1.59^{+0.56}_{-0.34}$\\ 
           &$-3.5\le {\rm [Fe/H]}<-1.3$ & 1,046 & $2.85^{+0.09}_{-0.08}$ & $1.52^{+0.59}_{-0.35}$\\ 
           &$-3.5\le {\rm [Fe/H]}<-1.4$ & 786   & $2.92^{+0.10}_{-0.10}$ & $1.84^{+1.59}_{-0.60}$\\ 
\hline
Velocity   &$-0.9\le {\rm [Fe/H]}<-0.8$ & 2,231 & $2.51^{+0.07}_{-0.07}$ & $1.40^{+0.25}_{-0.20}$\\ 
           &$-1.1\le {\rm [Fe/H]}<-0.9$ & 2,486 & $2.71^{+0.06}_{-0.06}$ & $1.64^{+0.58}_{-0.34}$\\ 
           &$-1.3\le {\rm [Fe/H]}<-0.8$ & 6,040 & $2.70^{+0.05}_{-0.05}$ & $1.40^{+0.18}_{-0.15}$\\ 
           &$-3.5\le {\rm [Fe/H]}<-1.1$ & 2,433 & $3.61^{+0.07}_{-0.07}$ & $1.30^{+0.36}_{-0.24}$\\ 
           &$-3.5\le {\rm [Fe/H]}<-1.3$ & 1,369 & $3.96^{+0.09}_{-0.09}$ & $1.37^{+0.48}_{-0.30}$\\ 
           &$-3.5\le {\rm [Fe/H]}<-1.4$ & 1,064 & $4.06^{+0.08}_{-0.08}$ & $1.61^{+0.97}_{-0.46}$\\ 
\hline
Action     &$-0.9\le {\rm [Fe/H]}<-0.8$ & 3,065 & $2.43^{+0.05}_{-0.05}$ & $1.65^{+0.28}_{-0.21}$\\ 
           &$-1.1\le {\rm [Fe/H]}<-0.9$ & 3,422 & $2.53^{+0.06}_{-0.06}$ & $2.37^{+0.47}_{-0.35}$\\ 
           &$-1.3\le {\rm [Fe/H]}<-0.8$ & 8,353 & $2.55^{+0.05}_{-0.05}$ & $1.80^{+0.17}_{-0.15}$\\ 
           &$-3.5\le {\rm [Fe/H]}<-1.1$ & 3,760 & $3.09^{+0.06}_{-0.06}$ & $1.74^{+0.64}_{-0.37}$\\ 
           &$-3.5\le {\rm [Fe/H]}<-1.3$ & 2,223 & $3.22^{+0.07}_{-0.07}$ & $1.83^{+0.77}_{-0.42}$\\ 
           &$-3.5\le {\rm [Fe/H]}<-1.4$ & 1,740 & $3.27^{+0.08}_{-0.08}$ & $2.14^{+2.25}_{-0.79}$\\ 
\hline
Velocity (45)  &$-1.2\le {\rm [Fe/H]}<-0.8$ & 5,794 & $2.68^{+0.05}_{-0.04}$ & $1.49^{+0.24}_{-0.19}$\\
           &$-1.5\le {\rm [Fe/H]}<-1.2$     & 1,108 & $3.44^{+0.09}_{-0.10}$ & $1.44^{+0.36}_{-0.24}$\\
           &$-1.8\le {\rm [Fe/H]}<-1.5$     & 467   & $4.11^{+0.12}_{-0.12}$ & $1.26^{+1.43}_{-0.48}$\\
           &$-3.5\le {\rm [Fe/H]}<-1.8$     & 447   & $4.25^{+0.09}_{-0.10}$ & $2.82^{+4.61}_{-1.61}$\\
           &$-0.9\le {\rm [Fe/H]}<-0.8$     & 2,288 & $2.55^{+0.06}_{-0.06}$ & $1.39^{+0.20}_{-0.16}$\\
           &$-1.1\le {\rm [Fe/H]}<-0.9$     & 2,617 & $2.74^{+0.06}_{-0.06}$ & $1.71^{+0.54}_{-0.33}$\\
           &$-3.5\le {\rm [Fe/H]}<-1.1$     & 2,688 & $3.71^{+0.06}_{-0.07}$ & $1.65^{+0.58}_{-0.33}$\\
\hline
\hline
Velocity (55)  &$-1.2\le {\rm [Fe/H]}<-0.8$ & 5,794 & $2.73^{+0.05}_{-0.05}$& $1.79^{+0.44}_{-0.29}$\\
           &$-1.5\le {\rm [Fe/H]}<-1.2$     & 1,108 & $3.56^{+0.10}_{-0.09}$ & $1.48^{+0.30}_{-0.22}$\\
           &$-1.8\le {\rm [Fe/H]}<-1.5$     & 467   & $4.34^{+0.10}_{-0.10}$ & $1.53^{+1.84}_{-0.61}$\\
           &$-3.5\le {\rm [Fe/H]}<-1.8$     & 447   & $4.42^{+0.10}_{-0.10}$ & $4.56^{+8.32}_{-4.56}$\\
           &$-0.9\le {\rm [Fe/H]}<-0.8$     & 2,386 & $2.59^{+0.06}_{-0.06}$ & $1.45^{+0.21}_{-0.17}$\\
           &$-1.1\le {\rm [Fe/H]}<-0.9$     & 2,647 & $2.79^{+0.06}_{-0.07}$ & $2.26^{+1.49}_{-0.63}$\\
           &$-3.5\le {\rm [Fe/H]}<-1.1$     & 2,821 & $3.88^{+0.07}_{-0.07}$ & $2.09^{+0.57}_{-0.36}$\\
\hline
\enddata
\tablecomments{We list results for various [Fe/H] bins for comparisons with other studies and sample sizes.}
\end{deluxetable}

\subsection{Correlation between the radial and vertical distances and metallicity}

\begin{figure}[htbp!]
\includegraphics[width =0.5\textwidth]{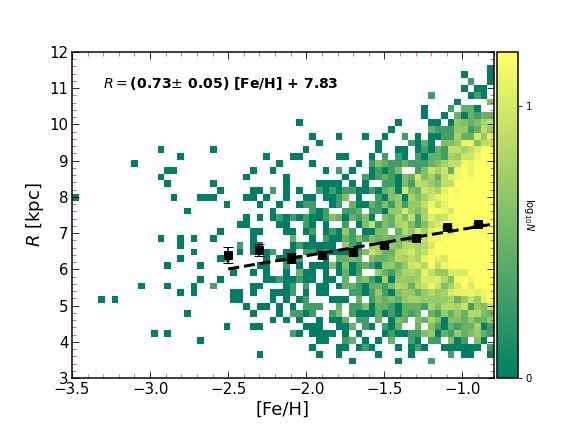}\\
\includegraphics[width =0.5\textwidth]{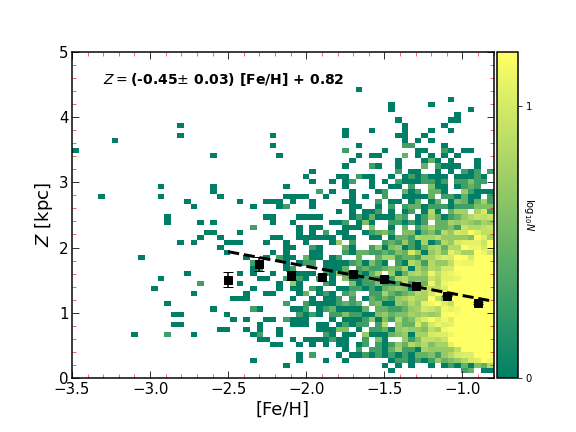}
\caption{Top: Radial metallicity gradient as a function of Galactocentric radial distance of the Atari disk sample. Bottom: Vertical metallicity gradient of the Atari disk as a function of \metal. Error bars denote the standard deviation in each bin and show the statistical uncertainty only.}
\label{fig:metallicity_gradients}
\end{figure}

The spatial-metallicity correlation of metal-poor stars in the Galactic disk places observational constraints on our understanding of the formation and evolution of the Milky Way system. The mean metallicity ($\langle\metal\rangle$) of stars at a particular region in the disk primarily depends on the gas accretion rate, the chemical composition of the early interstellar gas, and subsequent evolution of stars at that region. 
To investigate the presence of any correlation between metallicity and radial distance from the Galactic center ($R$), we apply a simple linear fit to the individual measurements of R vs. {\metal} in our Atari disk sample. 
The top panel of Figure~\ref{fig:metallicity_gradients} presents a 2d histogram of the $R$ distribution of the Atari disk sample as a function of \metal. The points and error bars represent the mean value and standard error of $R$ of bins of 0.20\,dex for visualization purposes. 
The slope of the dashed line represents a positive radial metallicity gradient (${\rm \pa R/\pa}\metal = 0.73\pm0.05$\,kpc ${\rm dex^{-1}}$). 
This result is different from what has been found for the canonical thick disk which is essentially flat. \citet{Recio2014} used 1,016 stars from the Gaia-ESO DR1 to chemically separate the disk components and found ${\rm \pa \metal/\pa R} = +0.006\pm0.008$ for the thick disk. \citet{Peng2018} used a kinematic approach to separate 10,520 stars taken from South Galactic Cap u-band Sky Survey and SDSS/SEGUE data and found ${\rm \pa \metal/\pa R} = -0.001\pm0.020$.
We note that the above studies have a higher metallicity range (\metal $\gtrsim -1.2$), caveating a direct comparison to our results.
Indeed, 

We also test for the presence of a {\metal} trend with absolute vertical distance from the Galactic plane ($|Z|$), using the same technique as for deriving the radial gradient, including using bin sizes of 0.2\,dex for visualization purposes. 
The bottom panel of Figure~\ref{fig:metallicity_gradients} shows the correlation between {\metal} and $|Z|$ and the best fit line, which represents a positive vertical metallicity gradient (${\rm \pa |Z|/\pa}\metal = -0.45\pm0.03$\,kpc ${\rm dex^{-1}}$). 
On average, the more metal-poor stars of the Atari disk are preferentially found at high $|Z|$ values. 
The relatively more metal-rich stars are on average located at $|Z| \lesssim 2$. 
Note that only stars within 5\,kpc were included in both of these analyses following \citet{Chiti2021_map}, to avoid selection effects toward more metal-poor stars at larger distances.

Following the last point, we investigated a few further avenues to assess to what extent selection effects affect our observed spatial-metallicity correlations. 
There are two primary ways in which selection effects could bias our gradients: (1) metal-poor stars are brighter than more metal-rich stars in the SkyMapper $v$ filter, making the most distant stars preferentially metal-poor; and (2) based on our exclusion of regions of high reddening in the initial sample \citep[see][for more details]{Chiti2021,Chiti2021_map}, there is an exclusion of low $Z$ stars at $R$ close to the galactic center; this could lead to an artificial $R$-{\metal} gradient given that lower metallicity stars are at high $Z$. 
The effect of (1) is generally accounted for by only considering stars within 5\,kpc from the Sun, which is the distance within which the SkyMapper filter should not significantly preferentially select metal-poor stars at large distances. 
To more stringently test this effect, we restrict our sample to stars within 4\,kpc and still find a positive $R$-{\metal} gradient. 
Restricting the sample to $<$2\,kpc results in no statistically significant gradient, but this is not necessarily surprising because we lose sensitivity to any gradient by restricting our sample to only nearby stars.
We investigate the effect of (2) by searching for a gradient in $R$-{\metal} at various $Z$ bins ($0.1 < Z < 1.1$ with 0.2\,kpc bins and $1.1 < Z < 3.0$ with 0.4\,kpc bins). 
In general, this analysis still leads to positive gradients at a given Z range, suggesting that (2) is not a significant effect. 
We note that the $R$-{\metal} gradient appears to not be significant at the lowest metallicities (below [Fe/H] $< -1.4$).

\subsection{Gradients with Rotational Velocity}
\begin{figure*}[htbp!]
\includegraphics[width =\textwidth]{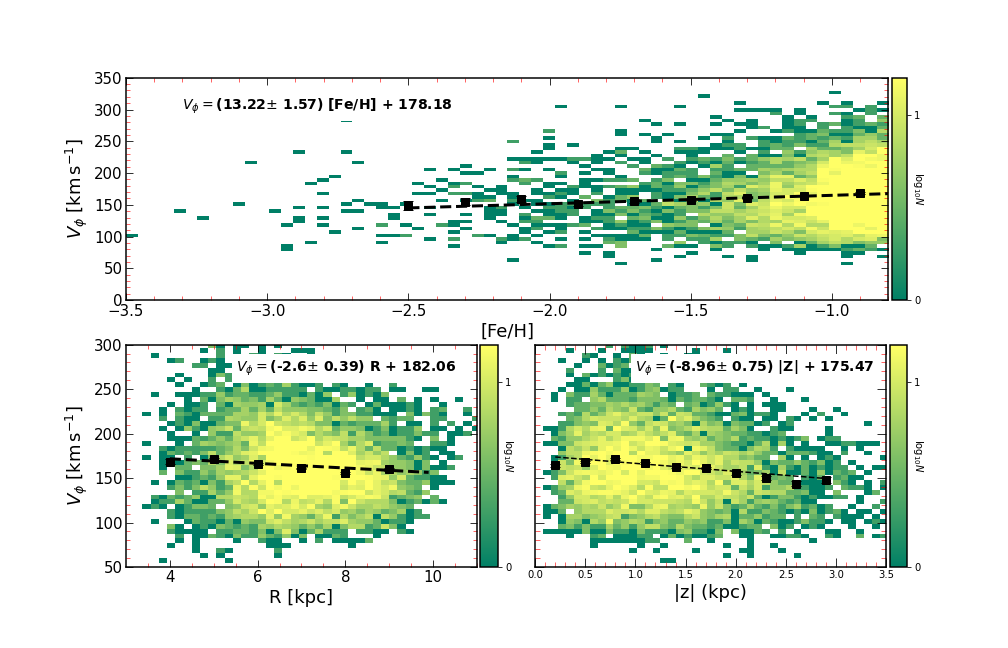}
\caption{Top: Rotational velocity as a function of {\metal} resulting in a gradient of ${\rm \pa V_{\phi}/\pa}\metal = 13.22\pm1.57$ km\,s$^{-1}$ ${\rm dex^{-1}}$. Bottom left: Rotational velocity as a function of the Galactocentric radial distance $R$ with a gradient of ${\rm \pa V_{\phi}/\pa}R = -2.6\pm0.4$ km\,s$^{-1}$ ${\rm kpc^{-1}}$. Bottom right: Rotational velocity as a function of scale height Z with a gradient of ${\rm \pa V_{\phi}/\pa}|z| = -8.96\pm0.75$ km\,s$^{-1}$ ${\rm kpc^{-1}}$. Error bars denote the standard deviations throughout.}
\label{fig:rotational_velocity}
\end{figure*}

We investigate the variation of rotational velocity $V_{\phi}$ versus {\metal}, $R$, and $|Z|$. 
The top panel of Figure~\ref{fig:rotational_velocity} shows a density plot of the rotational velocity versus the metallicity of stars in our Atari disk sample. 
The mean values and standard errors of $V_{\phi}$ in metallicity bins of 0.2\,dex are overplotted. There is an overall positive rotational velocity gradient as a function of {\metal} of ${\rm \pa V_{\phi}/\pa}\metal = 13.22\pm1.57$ km\,s$^{-1}$ ${\rm dex^{-1}}$. The lower left panel of Figure~\ref{fig:rotational_velocity} shows the rotational velocity gradient in the radial direction, ${\rm \pa V_{\phi}/\pa}R = -2.6\pm0.4$ km\,s$^{-1}$ ${\rm kpc^{-1}}$.
While the detection is statistically significant, the value of the gradient ($\sim2$\, km/s per 1 \,kpc) is small. There is also a negative correlation between $V_{\phi}$ and $|Z|$ of ${\rm \pa V_{\phi}/\pa}|z| = -8.96\pm0.75$ km\,s$^{-1}$ ${\rm kpc^{-1}}$.

Previous studies have found negative and positive slopes for the rotational velocity-metallicity gradient for the thin and thick disk populations, respectively. For reference, \citet{Lee2011} and \cite{Guiglion2015} used the chemical abundance approach to assign the stellar population membership for 17,277 and 7,800 stars, respectively. Using the thick disk samples in their studies, they reported rotational velocity gradients of ${\rm \pa V_{\phi}/\pa}\metal =+45.8\pm2.9$ km\,s$^{-1}$ ${\rm dex^{-1}}$ and ${\rm \pa V_{\phi}/\pa}\metal =+49\pm10$ km\,s$^{-1}$ ${\rm dex^{-1}}$, respectively. While \citet{Prieto2016} used 3621 APOGEE stars to measure ${\rm \pa V_{\phi}/\pa}\metal = -18 \pm 2$ km\,s$^{-1}$ ${\rm dex^{-1}}$. 
It is worth mentioning again that in all of these aforementioned studies, the metallicity range was \metal $> -1.0$ and so a direct comparison of results might not necessarily be accurate.

\subsection{Orbital Eccentricity}
\begin{figure*}[htbp!]
\includegraphics[width=\textwidth]{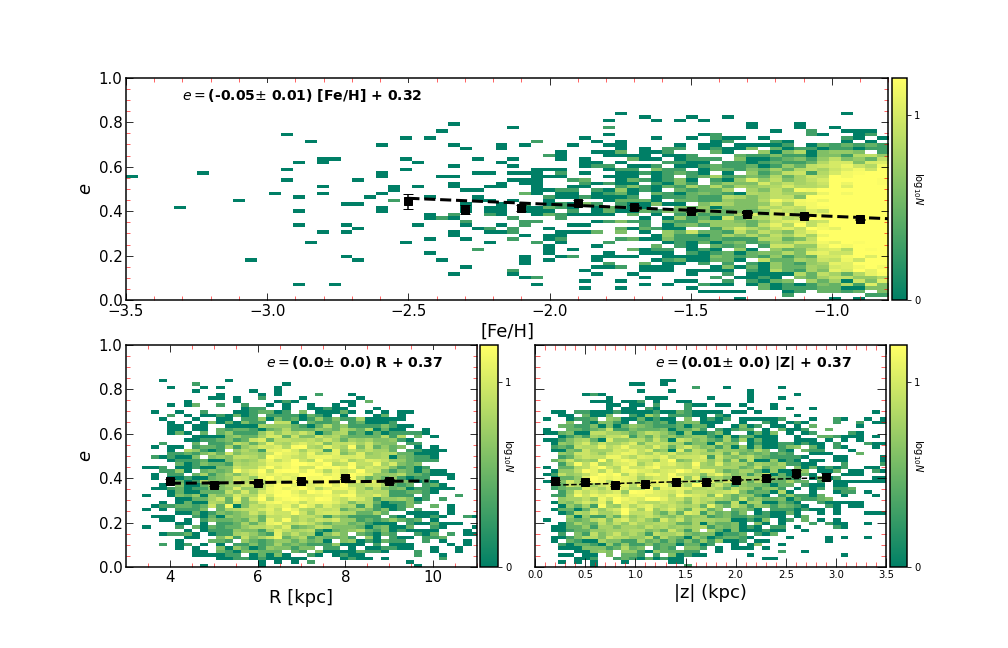}
\caption{Top: Orbital eccentricities as a function of {\metal} with a gradient of ${\rm \pa e/\pa}\metal = -0.05\pm0.01$ ${\rm dex^{-1}}$. Bottom left: $e$ as a function of the Galactocentric radial distance $R$ with a flat trend of ${\rm \pa e/\pa}R = 0.00\pm0.00$ ${\rm kpc^{-1}}$. Bottom right: $e$ as a function of scale height $Z$ with a positive gradient of ${\rm \pa e/\pa}|z| = +0.01\pm0.002$ ${\rm kpc^{-1}}$. Error bars denote the standard deviation throughout.}
\label{fig:orbital_eccentricities}
\end{figure*}

For our Atari disk sample, we investigate the relation between the orbital eccentricity, and {\metal}, $R$, and $|Z|$. Figure~\ref{fig:orbital_eccentricities} shows the observed trends of the orbital eccentricity as a function of {\metal}, $R$, and $|Z|$. The top panel in Figure~\ref{fig:orbital_eccentricities} shows eccentricity versus {\metal}. Results suggests that the orbital eccentricity increases as the metallicity decreases, with the most metal-poor star having fairly eccentric orbits. 
The best fit yields a slope of ${\rm \pa e/\pa}{\metal} = -0.05\pm0.01$\,dex$^{-1}$. The lower left panel presents an overall no significant correlation between the orbital eccentricity and $R$. The lower right panel identifies that the orbital eccentricity varies minorly with $|Z|$, with ${\rm \pa e/\pa}Z = +0.01\pm0.002$\,kpc$^{-1}$. 
In general, our Atari disk stars exhibit different orbital eccentricity with {\metal}, $R$, and $|Z|$ from the ones reported in the literature for the more metal-rich stars in the canonical thick disk \citep[see][figure 9]{Lee2011}.

\section{Comparisons with formation models}\label{sec:origin}
Compared to the thick and thin disks, the Atari disk has not been extensively studied or been regarded as a separate component of the Galactic disk until recently \citep{Carollo2019,An2020}. 
Accordingly, there are no detailed theoretical Atari disk formation scenarios discussed in the literature. In the absence of such, we will compare our observed characteristics of the Atari disk with predictions from the four main formation scenarios for a structurally distinct thick disk, as well as models with predictions regarding eccentricities, to gain insights into how the Atari disk formed and evolved.  
\subsection{Comparison with predictions of thick disk formation models based on \metal gradients}
\label{sec:scenarios}

We use the properties detailed in Section~\ref{sec:Atari_properties} to assess four main formation scenarios for a structurally distinct thick disk following \citet{Li2018} to learn about the origin of the Atari disk. We discuss each in detail below.

\textbf{1. Disk heating.} This scenario posits the dynamical heating of a pre-existing disk due to minor mergers. The disk will maintain its chemical or kinematic gradients \citep{Quinn1993, Kazantzidis2008} even after the merger(s). 
We observe a positive radial metallicity gradient and a negative vertical metallicity gradient for our Atari disk sample. 
We note that direct comparisons of the magnitude of our gradients are not possible to other studies in the literature due to the upper metallicity limit (\metal $< -0.75$) of our SMSS sample. However, our detection of a correlation between the radial distance and metallicity is not principally seen in some studies of this formation scenario of the thick disk, which disfavors this interpretation \citep[e.g.,][]{Recio2014,Peng2018}.

\textbf{2. Gas-rich merger.} At high redshifts, dwarf galaxies were likely all gas rich with few stars formed, including those that merged with the early Milky Way.
Any gas-rich deposit into the Milky Way's highly turbulent early disk would have expected to have triggered star formation \citep{Brook2004,Brook2007}. The subsequent stars that formed from this merger should likely show no obvious clumpy distribution in the integrals of motion space. 
Also, we would expect subsequent stars that formed to have formed in a the short timescale within the disk following a gas-rich merger, suggesting a flat metallicity behavior (no gradient) \citep{Cheng2012}. However, we do observe a gradient in our sample (see Figure~\ref{fig:metallicity_gradients}).
Thus, it is unlikely that the Atari population formed in a star formation episode after a gas-rich merger; although, it very likely could have been associated with the metal-poor stars that formed in an accreted galaxy before infall.

It is then interesting to consider the existence of significant numbers of metal-poor stars with $\metal < -2.5$ in this context. These stars do support accretion as the origin scenario of the Atari disk, as opposed to star formation following a gas-rich merger, which would lead to a population of higher metallicity stars with an average of $\metal \approx -0.6$. 
It would thus take a merger(s) to inject gas $\approx 1,000$ times more metal-poor to bring down the \metal\ of the disk's interstellar medium to allow the formation of such low-metallicity stars post-merger. 
This seems unlikely from having occurred.
However, such primitive stars may have easily formed in early low-mass systems which were accreted first by neighboring, more massive systems and eventually into the massive progenitor of the Atari disk.

\textbf{3. Direct accretion.} Cosmological simulations have shown that a direct accretion of dwarf-like galaxies coming in from specific directions can build up a thick disk by donating their content in a planar configuration \citep[for more details about such simulations, see][]{Abadi2003a,Abadi2003}. Either one major merger with a massive satellite or the accretion of a number of smaller systems would result in spatially distinct populations as measurable by differences in $h_{R}$ and $h_{Z}$ \citep[see][]{Gilmore2002}. Correlations between {\metal} and values for $h_{R}$ and $h_{Z}$ may indeed principally indicate multiple populations since such a scenario would deposit stars (not just gas) into the early Galactic disk that were formed within the progenitor system before the merger event. 
For example, an ex-situ Milky Way population would display a larger scale length compared to that of a population formed by an in-situ scenario \citep{Amores2017}. 
These stars (now present in the disk) would also still share similar integrals of motion. Finally, there would also be an expected observable metallicity gradient, both vertically and radially due to the different origins of the stars (accreted vs. in-situ formed stars). Eccentricities would also be distributed broadly and over a wider range \citep{Sales2009}.

As can be seen in Figure~\ref{fig:metallicity_gradients}, our Atari disk sample displays spatial metallicity gradients in both the vertical and radial direction. It also shows a broad range of eccentricities, as can be seen in Figure~\ref{fig:orbital_eccentricities}. We also find a moderate correlations of the scale length as a function of {\metal} for stars in the solar neighborhood vicinity, from $h_{R} \sim2.4$ to 2.9 (see Table~\ref{tab:lengths_and_heights}). The existence of this gradient aligns with predictions for an ex-situ population to have an increased scale length compared to an in-situ one \citep{Amores2017}, and supports the Atari disk to have an accretion origin. Unfortunately, the picture is less obvious regarding the behavior of the scale height. As our present Atari disk data is not suitable to draw strong evidence on the existence of a scale height metallicity gradient, we strongly recommend future studies with larger samples to further quantify this issue.

The direct accretion explanation is also supported by simulations of Milky Way-like galaxies in the \texttt{IllustrisTNG} simulation \citep{IllustrisTNG}. We analyze 198 simulated Milky Way analogs from \texttt{Illustris TNG50}, defined as disk galaxies with stellar masses of $M_* = 10^{10.5 - 11.2} M\odot$ in relative isolation at $z=0$ \citep[originally identified by][]{Engler21,Pillepich21}. Milky Way analogs are also reported in \texttt{Illustris TNG100} \cite[e.g.,][]{Mardini_2020}. When defining the thick disk of these galaxies, we look exclusively at star particles vertically located between 0.75 and 3\,kpc from the plane of the galaxy (excluding the area dominated by the thin disk or halo) and radially located between 3 and 15\,kpc from the center of the galaxy (excluding the area dominated by the bulge). We trace back the origin of the star particles in the thick disk at $z=0$ and find that the vast majority of stars were formed ex-situ: $95^{+3}_{-10}\%$ of the thick disk stars in each Milky Way analog have accretion origins. We also calculated {\metal} vs. radial distance gradients for the ex-situ thick disk populations. Among the 198 ``Milky Way thick disks", 71 of them have a positive [Fe/H] vs. radial distance gradient for their ex-situ population. This is 35\% of the simulated ex-situ thick disk populations. Several of the simulated ex situ thick disk populations have gradients that exactly match the Atari disk observations.

This percentage remains consistently high when we consider lower metallicity stars; for stars with \metal$< -0.8$, $96^{+3}_{-11}\%$ of the thick disk stars have accretion origins. This trend is supported by the results of \citet{Abadi2003}.

In summary, the behavior of the stellar spatial distributions, together with the eccentricity distribution, the scale length, and plausibly also the scale height lend support to a scenario in which the Atari disk formed by accretion event(s) similar to those studied in \citet{Abadi2003} as well as the \texttt{IllustrisTNG} simulation.

\textbf{4. Radial migration.} A radial migration scenario suggests that early dynamical interactions occurred when the metallicity of the interstellar medium was relatively low and the $\alpha$-abundance ratios of the disk stars were high. 
Specifically, interactions of the disk with the spiral arms can dilute the metallicity gradient by rearranging the orbital motion of stars \citep{Schonrich2009b}. This would lead to an exchange of thin and thick disk stars. Any outward migration places stars on higher orbits above or below the Galactic plane. By now, these migrated stars would have had enough time to also experience significant orbital mixing, thus contributing to the flattening of the gradient. 

Accordingly, only a small or even no correlation between the rotational velocity and the metallicity is expected in this scenario. Our sample does  show a significant correlation, as can be seen in Figure~ \ref{fig:rotational_velocity}, suggesting that radial migration has not played a major role in the (more recent) evolution of the Atari disk.\\

\subsection{Comparison with predictions of thick disk formation models based on orbital eccentricities}

In addition to metallicity gradients, information can also be gained from the distribution of the stellar orbital eccentricities \citep[e.g.,][]{Sales2009}. 
In the following, we consider eccentricity distribution predictions and compare them with our results in the context of the chemo-dynamic constraints already discussed in Section~\ref{sec:scenarios}. 
\citet{Sales2009} predict (their Figure~3) that \\
(i) a notable peak at low eccentricity should be present in the radial migration and gas-rich scenarios ($e \sim$0.2-0.3), \\
(ii) the accretion scenario has an eccentricity distribution that is broadly distributed but with a peak shifted towards higher values, and \\
(iii) the disk heating scenario has two peaks with the primary one being at $e \sim$ 0.2-0.3 and a secondary one located at $e \sim$ 0.8\footnote{This secondary peak is from the debris of an accreted/merged luminous satellite. If the disk heating is due to merging of subhalo(s), then this secondary peak might not likely exist.}.

\begin{figure}[htbp!]
\includegraphics[width =0.5\textwidth]{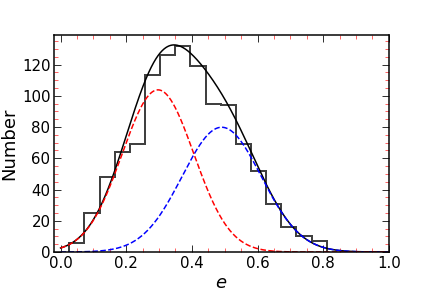}\\
\includegraphics[width =0.5\textwidth]{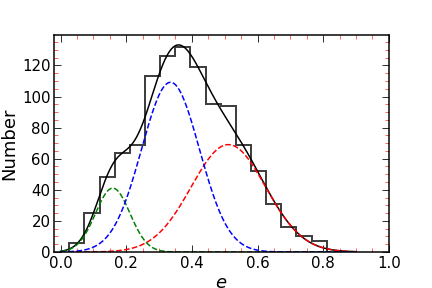}
\caption{Orbital eccentricity distribution for our Atari disk sample, using the scale length of $h_{R} = 2.48 \pm 0.15$\,kpc and scale height of $h_{Z} = 1.67 \pm 0.15$\,kpc in the ranges of $1 < |Z/h_{Z}| < 3$ and $2 < |R/h_{R}| < 3$, corresponding to figure~3 in \citet{Sales2009}. The solid black line represents the superposition of the individual Gaussians. Upper panel: best fit using two Gaussians with peaks at $e$ = 0.30 and 0.49. Lower panel:  best fit using three Gaussians with peaks at $e =$ 0.16, 0.33, and 0.51.}
\label{fig:eccentricities_distribution}
\end{figure}

In Figure~\ref{fig:eccentricities_distribution}, we show the orbital eccentricity distributions of our Atari disk sample and best-fitting Gaussians.
We can reproduce the observed distribution with two Gaussians in the upper panel (with peaks at $e = $0.33 and 0.54), and with three Gaussians in the lower panel (with peaks at $e = $0.24, 0.40, and 0.61). A significant number of stars with $e > 0.4$ in our sample principally argues against the importance of radial migration and these stars having originated from star-formation after the gas-rich scenario when considering the formation and evolution of the Atari disk. Instead, the presence of two or three broad, well separated peaks that fit the observed distribution quite well supports the prediction of the direct accretion model, in that the eccentricity distribution is quite broad.

We note that our distribution may qualitatively align with the disk heating scenario since we note a peak at $e\sim$0.3 and one at higher $e\sim0.6$.
However, the larger peak is not located quite as high as $e=0.8$ and the distribution can be well-described by more than two underlying gaussian distributions. 
Consequently, the disk heating scenario might play a role in the formation and evolution of the Atari disk, but the discussion in Section~\ref{sec:scenarios} and the overall broad eccentricity distribution suggest that an accretion scenario might be the dominant channel.\newline

\section{Findings and Conclusions}\label{sec:Conclusions}

Our detailed kinematic investigation of metal-poor stars selected from SkyMapper DR2 that are located in the Galactic disk has allowed us to identify the Atari disk and learn about its characteristics and speculate on its origin. 
In this Section, we synthesize our findings across Sections~\ref{sec:Atari_properties} and~\ref{sec:origin} and comment on other chemical characteristics of the Atari disk.

\subsection{Kinematic \& Spatial characterization of the Atari disk}

We have assessed and characterized the Atari disk with a new sample of 7,127 low-metallicity member stars and have outlined some of its properties in Section~\ref{sec:Atari_properties}. The main findings regarding the spatial distribution of the stars are as follows.

Our detailed study confirms earlier claims \citep{Carollo2019} of a notable velocity lag of the Atari disk compared to the canonical thick disk. The Atari disk has a well defined mean velocity of V$_{\phi} \approx 154$\,km\,s$^{-1}$ and FWHM = 63.9\,km\,s$^{-1}$, with individual values ranging from about 80 to 250\,km\,s$^{-1}$ as can be seen in Figure~\ref{fig:rotational_velocity}. 
A V$_\phi$ distribution with a distinct, net rotation characterizes that of a disk population, rather than a halo population. 
Our extensive kinematic selection results also align with previous findings \citep{Carollo2019} of a peak in angular momentum of L$_{z}\sim 1200$\,${\rm kpc}$\,km\,s$^{-1}$ when restricting our sample to stars with $R = 7$-$9$\,kpc from the Galactic center (due to L$_{z}$ increasing with increasing $R$ and assuming a constant rotational velocity). Correspondingly, other R brackets (R = 3-5, 5-7, 9-11\,kpc) have lower or higher L$_{z}$ values (see Figure~\ref{fig:lz_distribution}). 

The eccentricities of our Atari disk sample cover a broad range of values ranging from $e \sim 0.0$ to 1.0. The bulk of the stars have $e \sim 0.3$ to 0.7 which appears to be a range between that of the canonical thick disk and the Galactic halo.
A notable fraction of our stars have eccentricities different from typical canonical thick disk values (see Table~\ref{tab:populations}).
There is no significant sub-population of Atari disk stars with $e$ = 0.7-1 (only 61 stars), suggesting, again, that the Atari disk eccentricities range between typical thick disk and halo eccentricities. 

The velocity lag and the range of eccentricities offer strong support to the origin scenario in which the Atari disk forms as a result of a major accretion event in which a satellite (or satellites) plunged into the Galactic disk at early times while coming from a specific direction \citep{Abadi2003,Sales2009}. For comparison, a gas-rich merger is favored as the formation scenario for the canonical thick disk. This alone highlights distinct differences between the nature of these two populations.

An accretion scenario for the Atari disk may also principally be supported by a variable scale length and height with metallicity. 
However, investigating this point with Milky Way mass galaxies in the \texttt{IllustrisTNG} simulation \citep{IllustrisTNG} indicates that while accretion history does affect the scale height, other factors also play a role. Observationally, we do indeed find a small increase in scale length with decreasing metallicity from around 2.37\,kpc at $\metal\sim-1.0$ to nearly 3\,kpc at $\metal\sim-1.6$.

As discussed in Section~\ref{sec:Atari_properties}, the behavior of the scale height with decreasing metallicity is somewhat inconclusive due to significant uncertainties (arising from small sample sizes and difficulties in measuring the first term in Equation~\ref{eq:Jeans4}). Therefore, we strongly recommend future studies of larger Atari disk data attempting to investigate the existence of a $h_{z}$ gradient with \metal.

\subsection{Very and extremely metal-poor stars in the Atari disk as identified in our SMSS sample}

The \metal\ behavior of the Atari disk appears to be significantly different from that of the canonical thick disk, as it stretches to much lower \metal, not unlike what is canonically found in the (inner and outer) halo populations.

We searched our Atari disk sample for low-metallicity stars and identified 261 stars with $\metal <-2.0$ (4\,\% of our sample), 55 stars with $\metal <-2.5$ (1\,\% of our sample), and 7 stars with $\metal <-3.0$ (0.1\,\% of our sample). 
We list stars with $\metal <-2.5$ in Table~\ref{tab:SMSS_EMP}, along with any available literature metallicities.
To check again whether these stars could be halo stars, we inspected the long-term orbital histories ($Z_{max}$ and orbital eccentricity) of these objects. The bulk of these stars have $Z_{max} < 3$\,kpc, suggesting that they are indeed not part of the halo population. Any stars with a higher $Z_{max}$ stars appear not have eccentricities exceeding e $\sim 0.6$, again confirming no halo membership. 
This leaves the questions whether our sample would contain any thick disk stars. 
However, we find these low-metallicity stars to generally have either too high an eccentricity or $Z_{max}$ values to be associated with the thick disk (as shown in Table~\ref{tab:populations}). 

Of our 55 Atari disk stars with $\metal <-2.5$, $\sim60$\% are readily found in the Simbad database \citep{SIMBAD2000}. 
Table~\ref{tab:SMSS_EMP} lists the literature metallicities and corresponding references. 
We note that four stars have also previously been classified to have disk-type kinematics, as is noted in Table~\ref{tab:SMSS_EMP}. 
Overall, for these stars with $\metal <-2.5$ (excluding those with measurements from the GALAH survey), the photometric {\metal} estimates from \citet{Chiti2021} agree very reasonably with those from the literature, with a mean difference of $0.11\,\pm\,0.05$. 

Several re-discovered stars display interesting chemical abundance patterns. 
Five stars are limited-$r$ stars with light neutron-capture element enhancements \citep{Frebel2018_Review,Hansen2018RPA}, two stars are mildly $r$-process enhanced \citep{Hansen2018RPA, Ezzeddine2020_RPA} and two are carbon-enhanced metal-poor (CEMP) stars. Of the seven stars with \metal $<-3.0$, two are already known in the literature. One was analyzed by the R-Process Alliance \citep{Sakari2018_RPA} and found to be a CEMP star, and the other was studied by \citep{Best_and_Brightest}. 
Of the remaining five, we have observed one star and chemical abundance results will be reported in X. Ou et al. (in prep.). 

We also decided to search the two original parent samples (the individual action and velocity-based selection samples) for additional metal-poor stars. 
We find eleven and seven more very and extremely metal-poor stars in the sample, respectively. 
Of those extra eleven action-selected stars, eight stars appear to have likely halo kinematics (Z$_{max} > 3.0$\,kpc and/or $e > 0.7$). Five have Z$_{max}<4$ but large eccentricities ($e>0.65$). 
We add the remaining three stars to our Atari disk sample. 
Of the seven velocity-selected stars, six have likely halo kinematics (Z$_{max} > 3.0$\,kpc and/or $e > 0.7$). We add only the remaining one to our sample. 
The stars are listed in Table~\ref{tab:SMSS_EMP}.

While these four metal-poor stars were not selected into our final common sample with the highest likelihood for being the most representative of Atari disk stars, we note that it is highly likely that they do belong to the Atari disk. They clearly do not belong to \textit{another} population, as per their kinematic properties. Hence, when searching for the most metal-poor stars it is critical to consider these two selection methods individually as well, given the rarity of such stars.

The existence of large numbers of bona-fide very and extremely metal-poor stars in the Atari disk significantly supports an accretion scenario. A massive progenitor system must have undergone at least a limited amount of (early) chemical evolution that produced an early population of low-metallicity stars. For comparison, the thick disk is not known for having many such metal-poor stars, and the gas-rich merger scenario would not support the existence of a large fraction either.
We discuss possible scenarios for the nature of the potential progenitor further below. 

\begin{deluxetable*}{lrlrrrrr}
\tiny
\caption{Very and extremely metal-poor SMSS stars with Atari disk kinematics}
\label{tab:SMSS_EMP}
\tablehead{
\colhead{R.A. (J2000)} & 
\colhead{Decl. (J2000)} & 
\colhead{Gaia ID} & 
\colhead{\metal$_{phot}$} & 
\colhead{\metal$_{lit}$} & 
\colhead{Z$_{max}$ [kpc]} &
\colhead{e}&
\colhead{References for \metal$_{lit}$} }
\startdata
12 35 57.41&$-$34 40 19.70&6158268802159564544&$-$3.00&\nodata& 4.10&0.45&\\
06 51 52.43&$-$25 07 02.22 & 2921777156667244032\tablenotemark{b} & $-$3.02 & \nodata & 0.89 & 0.21 &  \\
20 05 28.77&$-$54 31 25.97&6473118900280458240&$-$3.04&$-$3.01& 4.10&0.18&\cite{Best_and_Brightest}\\
16 23 29.34 &$-$65 17 53.33 & 5827787046034460672\tablenotemark{b} & $-$3.04 & $-$2.31 & 3.99 & 0.31 & \cite{Buder+2021}\\
09 29 49.73&$-$29 05 59.03&5633365176579363584&$-$3.10&$-$2.88& 0.90&0.45&\cite{Sakari2018_RPA}\\
19 25 0.04 & $-$15 57 43.35 & 4181010754108521472\tablenotemark{c} & $-$3.14 & \nodata & 1.49 & 0.28  & \\
18 11 39.41&$-$46 45 43.50&6707545022835614464&$-$3.24&\nodata& 1.90&0.42&\\
21 48 07.05&$-$43 43 23.74&6565897654232474240&$-$3.24&\nodata& 4.80&0.57&\\
14 59 14.59&$-$10 49 42.34&6313811313365806208&$-$3.31&\nodata& 3.70&0.41&\\
23 30 19.62&$-$08 13 15.20&2438343952886623744&$-$3.48&$-$3.20& 4.20&0.56&X. Ou et al 2022 (in prep)   \\
15 45 12.76 &$-$31 05 29.22 & 6016676026902257280\tablenotemark{b} & $-$3.53 & \nodata & 2.10 & 0.22 & \\
\enddata
\tablecomments{A short version is shown here to illustrate the table form and content, but the full content is accessible in the online table.}
\end{deluxetable*}

\begin{deluxetable*}{lrlrrrrrr}
\caption{Very and extremely metal-poor stars with Atari disk kinematics from the literature (collected through JINAbase)}
\label{tab:jina30}
\tablehead{
\colhead{R.A. (J2000)} & 
\colhead{Decl. (J2000)} & 
\colhead{Simbad identifier} & 
\colhead{\metal$_{lit}$} & 
\colhead{\cfe} & 
\colhead{Z$_{max}$\,[kpc]} & 
\colhead{e} &
\colhead{Ref} }
\startdata
\hline\hline
16 28 56.15&$-$10 14 57.10&2MASS J16285613$-$1014576 &$-$2.00&0.24   &0.58&0.43&\cite{Sakari2018_RPA}\\
18 28 43.44&$-$84 41 34.81&2MASS J18284356$-$8441346 &$-$2.03&$-$0.39&3.19&0.32&\cite{Ezzeddine2020_RPA}\\
05 52 15.78&$-$39 53 18.47&TYC 7602-1143-1       &$-$2.05&\nodata&1.33&0.69&\cite{Ruchti2011}\\
00 25 50.30&$-$48 08 27.07&2MASS J00255030$-$4808270 &$-$2.06&0.27   &1.61&0.53&\cite{Barklem2005}\\
14 10 15.84&$-$03 43 55.20&2MASS J14101587$-$0343553 &$-$2.06&$-$0.09&1.17&0.71&\cite{Sakari2018_RPA}\\
02 21 55.60&$-$54 10 14.40&2MASS J02215557$-$5410143 &$-$2.09&0.00   &3.16&0.33&\cite{Holmbeck+2020_rpa}\\
23 05 50.54&$-$25 57 22.29&CD $-$26$^{\circ}$ 16470  &$-$2.13&\nodata&2.75&0.34&\cite{Ruchti2011}\\
13 43 26.70&$+$15 34 31.10&HD119516                 &$-$2.16&\nodata&1.07&0.52&\cite{For_2010}\\
15 54 27.29&$+$00 21 36.90&2MASS J15542729+0021368   &$-$2.18&0.42   &0.88&0.51&\cite{Sakari2018_RPA}\\
03 46 45.72&$-$30 51 13.32&HD23798                  &$-$2.22&\nodata&0.83&0.54&\cite{Roederer_2010}\\
15 02 38.50&$-$46 02 06.60&2MASS J15023852$-$4602066 &$-$2.23&$-$0.16&0.97&0.30&\cite{Holmbeck+2020_rpa}\\
04 01 49.00&$-$37 57 53.40&2MASS J04014897$-$3757533 &$-$2.28&$-$0.30&2.61&0.52&\cite{Holmbeck+2020_rpa}\\
23 02 15.75&$-$33 51 11.03&2MASS J23021574$-$3351110 &$-$2.29&   0.37&2.08&0.44&\cite{Barklem2005}\\
11 41 08.90&$-$45 35 28.00&2MASS J11410885$-$4535283 &$-$2.32&\nodata&1.33&0.63&\cite{Ruchti2011}\\
09 29 49.74&$-$29 05 59.20&2MASS J09294972$-$2905589 &$-$2.32&   0.11&0.92&0.46&\cite{Sakari2018_RPA}\\
19 16 18.20&$-$55 44 45.40&2MASS J19161821$-$5544454 &$-$2.35&$-$0.80&2.37&0.49&\cite{Hansen2018RPA}\\
22 49 23.56&$-$21 30 29.50&TYC 6393-564-1        &$-$2.38&\nodata&2.96&0.54&\cite{Ruchti2011}\\
00 31 16.91&$-$16 47 40.79&HD2796                   &$-$2.40&$-$0.48&1.13&0.68&\cite{Mardini_2019a}\\
11 58 01.28&$-$15 22 18.00&2MASS J11580127$-$1522179 &$-$2.41&   0.62&3.62&0.38&\cite{Sakari2018_RPA}\\
15 14 18.90&$+$07 27 02.80&2MASS J15141890+0727028   &$-$2.42&   0.47&3.66&0.55&\cite{Roederer_2010}\\
05 10 35.47&$-$15 51 38.30&UCAC4 371-007255        &$-$2.43&\nodata&0.74&0.60&\cite{Cohen_2013}\\
16 10 31.10&$+$10 03 05.60&2MASS J16103106+1003055   &$-$2.43&   0.53&2.49&0.23&\cite{Hansen2018RPA}\\
05 51 42.14&$-$33 27 33.76&TYC 7062-1120-1       &$-$2.46&\nodata&1.71&0.68&\cite{Holmbeck+2020_rpa}\\
23 16 30.80&$-$35 34 35.90&BPS CS30493$-$0071       &$-$2.46&$-$0.01&1.26&0.40&\cite{Roederer2014}\\
01 07 31.23&$-$21 46 06.50&UCAC4 342-001270        &$-$2.55&\nodata&3.28&0.11&\cite{Cohen_2013}\\
18 36 23.20&$-$64 28 12.50&2MASS J18362318$-$6428124 &$-$2.57&   0.10&1.37&0.24&\cite{Hansen2018RPA}\\
18 40 59.85&$-$48 41 35.30&2MASS J18405985$-$4841353 &$-$2.58&   0.60&1.50&0.48&\cite{Ezzeddine2020_RPA}\\
18 36 12.12&$-$73 33 44.17&2MASS J18361214$-$7333443 &$-$2.61&   0.10&2.13&0.48&\cite{Ezzeddine2020_RPA}\\
01 49 07.94&$-$49 11 43.16&CD $-$49$^{\circ}$ 506    &$-$2.65&\nodata&2.75&0.49&\cite{Ruchti2011}\\
09 47 19.20&$-$41 27 04.00&2MASS J09471921$-$4127042 &$-$2.67&$-$0.42&2.40&0.35&\cite{Holmbeck+2020_rpa}\\
21 51 45.74&$-$37 52 30.88&2MASS J21514574$-$3752308 &$-$2.76&   0.38&1.87&0.35&\cite{Beers_1992}\\
22 24 00.14&$-$42 35 16.05&2MASS J22240014$-$4235160 &$-$2.77&   0.14&2.45&0.49&\cite{Roederer2014}\\
15 56 28.74&$-$16 55 33.40&SMSS J155628.74$-$165533.4&$-$2.79&   0.36&2.15&0.36&\cite{Jacobson_2015}\\
22 02 16.36&$-$05 36 48.40&2MASS J22021636$-$0536483 &$-$2.80&$-$0.25&3.34&0.58&\cite{Hansen2018RPA}\\
03 01 00.70&$+$06 16 31.87&BPS CS31079$-$0028       &$-$2.84&\nodata&1.34&0.62&\cite{Roederer_2010}\\
04 19 45.54&$-$36 51 35.92&2MASS J04194553$-$3651359 &$-$2.89&   0.06&3.58&0.52&\cite{Roederer2014}\\
21 20 28.65&$-$20 46 22.90&BPS CS29506$-$0007       &$-$2.94&\nodata&1.02&0.45&\cite{Roederer2014}\\
14 35 58.50&$-$07 19 26.50&2MASS J14355850$-$0719265 &$-$2.99&$-$0.40&3.02&0.45&\cite{Ezzeddine2020_RPA}\\
20 42 48.77&$-$20 00 39.37&BD $-$20$^{\circ}$ 6008   &$-$3.05&$-$0.43&1.40&0.32&\cite{Roederer2014}\\
06 30 55.57&$+$25 52 43.81&SDSS J063055.57+255243.7\tablenotemark{a}&$-$3.05&\nodata&0.74&0.50&\cite{Aoki_2013}\\
13 03 29.48&$+$33 51 09.14&2MASS J12222802+3411318&$-$3.05&$-$0.18&3.93&0.61&\cite{Lai_2008}\\
00 20 16.20&$-$43 30 18.00&UCAC4 233-000355&$-$3.07&3.02&2.70&0.52&\cite{Cohen_2013}\\
13 19 47.00&$-$04 23 10.25&TYC 4961-1053-1\tablenotemark{a}&$-$3.10&$-$0.52&4.50&0.30&\cite{Hollek_2011}\\
12 45 02.68&$-$07 38 46.95&SDSS J124502.68$-$073847.0\tablenotemark{a}&$-$3.17&2.54&4.00&0.50&\cite{Aoki_2013}\\
14 16 04.71&$-$20 08 54.08&2MASS J14160471$-$208540&$-$3.20&1.44&1.92&0.54&\cite{Barklem2005}\\
13 22 35.36&$+$00 22 32.60&UCAC4 452-052732&$-$3.38&\nodata&2.98&0.45&\cite{Cohen_2013}\\
23 21 21.56&$-$16 05 05.65&HE 2318$-$1621\tablenotemark{a}&$-$3.67&0.54&3.64&0.58&\cite{Placco_2020}\\
11 18 35.88&$-$06 50 45.02&TYC 4928-1438-1\tablenotemark{a}&$-$3.73&0.08&4.71&0.31&\cite{Hollek_2011}\\
13 02 56.24&$+$01 41 52.12&UCAC4 459-050836&$-$3.88&1.34&2.80&0.26&\cite{Barklem2005}\\
09 47 50.70&$-$14 49 07.00&HE 0945$-$1435\tablenotemark{a}&$-$3.90&$<$2.03&0.76&0.57&\cite{Hansen_2015ApJ}\\
10 55 19.28&$+$23 22 34.02& SDSS J105519.28+232234.0\tablenotemark{a}&$-$4.00 & $<$0.70 & 2.20 & 0.45&\cite{Aguado_2017} \\
14 26 40.33&$-$02 54 27.49& HE 1424$-$0241\tablenotemark{a}&$-$4.05 & $<$0.63 & 3.50 & 0.41 &\cite{Cohen_2007ApJ} \\
12 47 19.47&$-$03 41 52.50& SDSS J124719.46$-$034152.4 &$-$4.11 & $<$1.61 & 1.71 & 0.24& \cite{Caffau2013} \\
12 04 41.39&$+$12 01 11.52& SDSS J120441.38+120111.5\tablenotemark{a}&$-$4.34 & $<$1.45 & 3.35 & 0.39& \cite{Placco2015ApJ} \\ 
10 29 15.15&$+$17 29 27.88& SDSS J102915.14+172927.9&$-$4.99 & $<$0.70 & 2.47 & 0.06&\cite{Caffau_2011C} \\
\enddata
\tablenotetext{a}{Stars that have too large uncertainties in their Gaia EDR3 astrometric data to be useful.
}
\end{deluxetable*} 
\subsection{Very and extremely metal-poor stars in the Atari disk as identified in literature samples}\label{sec:JINAbase}

Knowing about the existence of very and extremely metal-poor stars in the Atari disk, we also applied our selection procedure to known samples of metal-poor stars to identify additional ones. The SMSS sample used in the analysis presented here does not cover e.g., Northern hemisphere stars, warm low-metallicity stars, extremely metal-poor stars with $\metal<-3.5$, and very faint stars. This leaves room for more discoveries. 

Hence, we chose to investigate the entire data set compiled in the JINAbase \citep{jinabase}. The latest version is publicly available on \href{https://github.com/Mohammad-Mardini/JINAbase}{GitHub}\footnote{https://github.com/Mohammad-Mardini/JINAbase}. We cross-matched all stars with \metal $<-2.0$ (2,302 stars) of the JINAbase catalog with \textit{Gaia} EDR3 and applied the same quality cuts (\texttt{astrometric\_excess\_noise} $<$\,1\,$\mu$as and Parallax$\_$over$\_$error $\geqslant$\,5). We then collected radial velocities for these stars if they were not already listed in JINAbase. This resulted in a sample of 1,098 stars from which we identified a total of 47 Atari disk stars (5\%) with $\metal<-2.0$. Of those 22 stars have $\metal<-2.5$, eight have $\metal<-3.0$, and two have $\metal<-4.0$. A number of stars show interesting chemical abundance features. Table~\ref{tab:jina30} lists all these Atari disk stars. These stars have a mean $<V_{\phi}> = 152$\,km\,s$^{-1}$ which is highly consistent with our full SMSS Atari disk sample's mean velocity of 154\,km\,s$^{-1}$ (see Table~\ref{tab:populations}). Eccentricities range from about 0.05 to 0.7, with typical values around 0.4 to 0.5.

During our investigation of the literature sample as collected from JINAbase, we noticed a number of e.g. faint stars with low proper motion uncertainties and high uncertain parallaxes, leading to their exclusion based on the adopted Gaia astrometry quality cut. In order to prevent the discovery of additional exremely metal-poor stars due insufficient data quality, we thus opted to do an additional probability analysis to identify any potential Atari disk members. 
We drew 10,000 realizations of each of the 6-d astrometries for these JINAbase stars with \metal $< -3$ assuming a normal distribution. We then rerun the whole analysis using these possible combinations to determine the most likely membership for a halo, thin disk and thick-disk-like kinematic behavior. We only identified nine stars with \metal\,$\sim-3.0$. Of those, three stars have $\metal\lesssim-4.0$. 
We also identified three stars with \metal\,$<-4.0$ with a 50\% probability for being be part of the Atari disk. However, upon further inspection, we find both their Z$_{max}$ and eccentricities too high to be bona-fide Atari disk stars. We thus do not include them in our sample.

\subsection{The metallicity distribution function of the Atari disk}

\begin{figure}[htbp!]
\includegraphics[width =0.5\textwidth]{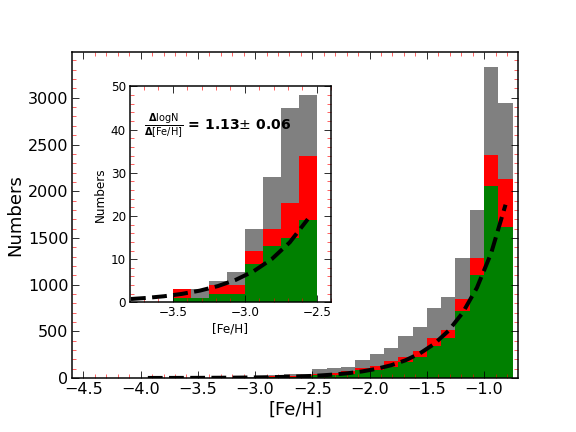}\\
\includegraphics[width =0.5\textwidth]{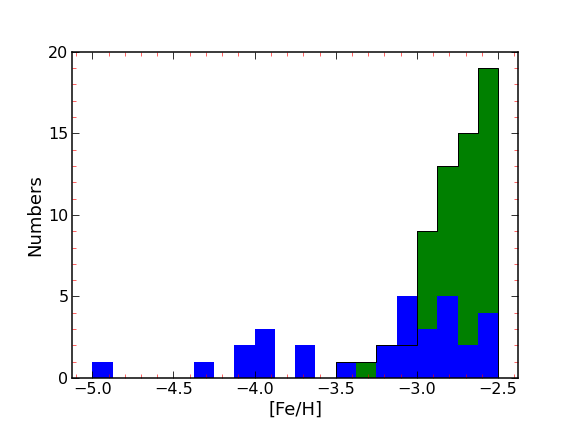}
\caption{Top: Metallicity distribution function (MDF) for stars with \metal $< -0.8$ of the action-selected method (gray histogram), the velocity-selected method (red histogram), and our Atari disk sample (green histogram). The inset figure shows the MDFs of stars with \metal $<-2.5$ and the best-fitting exponential ($\Delta\log {\rm N} / \Delta \metal = 1.13 \pm 0.06$). Bottom: MDF of the metal-poor tail (with \metal $< -2.5$) of our Atari disk sample (green histogram) compared with the sample of stars that we identify as Atari disk stars from JINAbase (blue histogram).}
\label{fig:Atari_MDF}
\end{figure}

The upper panel of Figure~\ref{fig:Atari_MDF} shows the metallicity distribution function for our final Atari disk sample (green histogram), and also for the velocity-selection (red histogram) and action-selection methods (gray histogram). 
The distributions look very similar, albeit the overall numbers are different. 
Our main sample shows an exponential decrease in stars with decreasing \metal, but with stars reaching down to \metal\,$\sim-3.5$, unlike what has been found for the canonical thick disk. 
The distribution of the two parent samples support this overall behavior.

The inset in Figure~\ref{fig:Atari_MDF} shows just the metal-poor tail ({\metal} $< -2.5$) of the MDFs, with a best-fitting exponential ($\Delta\log {\rm N} / \Delta \metal = 1.13 \pm 0.06$). The best-fitting exponential curve (dashed black line) drops to zero at $\metal \approx -4.0$, supporting the existence of only a handful of Atari disk stars (as identified in the literature) with $\metal \approx -4.0$ (see Table~\ref{tab:jina30}). 

The lower panel of Figure~\ref{fig:Atari_MDF} then shows the very metal-poor tail (\metal $< -2.5$) of our Atari disk sample (green histogram) in comparison with the stars from the literature (blue histograms) that we identified as Atari disk stars (see Section~\ref{sec:JINAbase}). Both samples show that Atari disk contains a number of stars with \metal $<-3.0$, with the literature sample containing stars with $\metal < -4.0$ (in agreement with our best-fitting exponential curve) and even $\metal \approx -5.0$.

The MDF currently shows no clear peak up to  $\metal < -0.8$. However, there is likely an increasing member contamination by canonical thick disk stars as [Fe/H] becomes higher than $\sim-1.0$. 
Assuming the upper bound of the mean metallicity of the Atari disk is set by [Fe/H] = $-0.8$ (following from the simple selection recipe presented in and previous Atari disk selections in \citealt{Beers+2014} and \citealt{Naidu2020}), 
we estimate a conservative upper limit to the stellar mass of the progenitor system of $\sim$10$^{9}$\,M$_{\odot}$ from the mass-metallicity relation in \citet{Kirby2013}. 
The progenitor mass is likely much lower than this value, though, as the Atari disk ought to be dwarfed by the thick disk (which has a mass of 1.17 $\times 10^{10}$\,M$_{\odot}$) since the Atari disk kinematic signature is only detectable relative to the thick disk in the low metallicity regime.

\subsection{Chemical abundance characteristics of the Atari disk}

An accretion origin of the Atari disk would imply that distinct chemical abundance signatures may be identifiable among Atari disk stars, in particular at the lowest metallicities. We briefly comment on findings as obtained from the available literature abundances. A more complete discussion will be presented in X. Ou et al. (2022, in prep.).

Besides the fact that a significant number of stars with \metal $<-3.0$ seem to belong to the Atari disk, we find several interesting chemical signatures. 
The 56 (47 + 9 stars with low quality parallaxes) JINAbase-selected stars with $\metal<-2.0$ display an average [$\alpha$/Fe] $\geqslant 0.3 $\,dex, which is a feature of enrichment by core-collapse supernovae and is generally seen in more metal-poor stars. 
This enhanced $\alpha$-abundance behavior is unlike that of the thick disk, which generally shows [$\alpha$/Fe] $\lesssim 0.2$\,dex.
However, we note that the lower [$\alpha$/Fe] of the thick disk may be due to its higher metallicity range than the Atari disk.

Carbon enhancement among metal-poor stars is regarded as a signature of very early star formation and commonly found among the most metal-poor halo stars (e.g., \citealt{Frebel2015_Review}). 
Of the 17 stars with \metal\,$\lesssim-3.0$, three are CEMP stars with \cfe $>0.7$ (17\%). 
If we were to also count the two additional stars with upper limits that do not exclude a CEMP-nature, the fraction would increase to 29\%. 
Interestingly, none of the five stars with \metal\,$\lesssim-4.0$ appear to be carbon-enhanced at face value, although the two stars with upper limits on carbon are in this metallicity group. If they were indeed CEMP stars, the fraction of CEMP stars could be as high as 41\%.
For comparison, in the halo, 24\% of stars with \metal\,$<-2.5$ are CEMP stars, 43\% at \metal\,$<-3.0$, 60\% at \metal\,$<-3.5$, and 81\% at \metal\,$<-4.0$ \citep{Placco2014_CEMP}. It thus remains to be seen what the CEMP fraction is among the lowest metallicity stars in the Atari disk but the existence of at least three CEMP stars with \metal\,$\lesssim-3.0$ points to Population\,III inhomogeneous, faint supernova-driven enrichment \citep{Umeda_and_Nomoto_2003} within the earliest star forming systems which offers additional support for an accretion origin of the Atari disk.

\subsection{On the origin and history of the Atari disk}

Overall, we find that the Atari disk is principally \mbox{disk-y} in nature as it appears to be confined to a somewhat puffed up, disk-like structure. This is illustrated by the fact that the long term orbital evolution of Atari disk stars, including the most metal-poor ones, shows them to remain within Z$_{\text{max}}$ = 3\,kpc. The Atari disk appears to be distinct from the canonical thick disk due to its rotational velocity lagging by $\sim30$\,km\,s$^{-1}$ and its distinct peak in angular momentum at a given radius. Moreover, Atari disk stars exhibit more varied eccentricities than the canonical thick disk and the Atari disk stellar population exhibits a significant low-metallicity tail. 

Based on our discussion in Section~\ref{sec:origin}, the origin of the Atari disk likely stems from an early accretion event. 
However, going forward, it will be important to compare our findings with results from tailored theoretical formation models for the Atari disk. In particular, cosmological simulations focusing on disk formation will be able shed more light on how multiple disk components (e.g., the Atari disk) may form and evolve and the nature of any transitory stellar populations between the components. 
We isolated our sample to \metal $< -0.8$ to preferentially include Atari disk stars, but future observational investigations may be able to remove this metallicity criterion if a sufficiently pure dynamical criterion were established. 
An improved selection criterion is particularly noteworthy, as an accretion scenario may support the existence of higher {\metal} stars depending on the chemical evolution of the progenitor system. 

\begin{figure}[htbp!]
\includegraphics[width =0.5\textwidth]{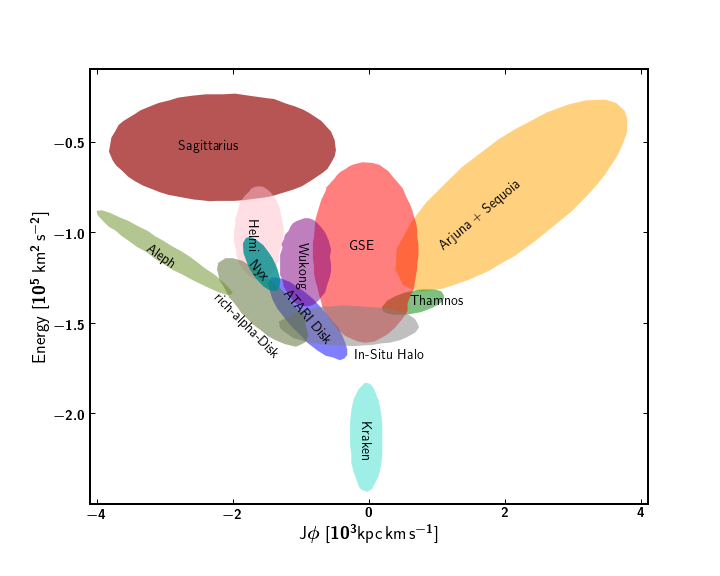}
\caption{Schematic approximate location of various galactic structures in the $E-L_{z}$ space adopted from \citet{Naidu2020}. The approximate location of our Atari disk sample is highlighted by the blue ellipsoid. Our Atari disk sample overlaps with the $E-L_{z}$ space of the Wukong, the in-situ halo, and the $\alpha$-rich disk (canonical thick disk).}
\label{fig:E_Lz_space}
\end{figure}

To investigate whether any currently known accreted structures could feasibly be related to the Atari disk, we compared the kinematic properties of the Atari disk to those of several recently identified structures (see Figure~\ref{fig:E_Lz_space}).
We list some comparisons below:

\textit{Gaia-Sausage-Enceladus}:
The Gaia-Sausage-Enceladus (GSE) was identified using varied selection methods \cite[e.g.,][]{GSE_Belokurov2018, Helmi_review2020}, which result in differing degrees of contamination with overlapping structures. 
However, the GSE stars cover a narrow range in rotational velocity centered at $\langle V_{\phi}\rangle = 0$\,km\,s$^{-1}$ coupled with a broad $V_{r}$ distribution. 
The orbital eccentricities typically are $e > 0.8$, and the GSE has a narrow MDF that peaks at \metal $\approx -1.17$ \citep{Feuillet2020}. 
A comparison of these properties with kinematics properties of our Atari disk sample readily shows differences, suggesting no association with the GSE structure.  

\textit{Kraken}:
The Kraken is the largest ($2 \times 10^{8}$ M$_{\odot}$) and oldest ($\approx 11$\,Gyr) galactic merger in the Milky Way's history, as described in \citet{Kraken2020}.
The spatial boundaries of the Kraken remnants are not well constrained. 
However, field stars originating from massive satellites reside deeper in the gravitational potential of the Milky Way with a clear separation due to dynamical friction \citep[][]{Amorisco2017}.
This suggests that the Kraken's debris would settle in low-energy and more eccentric orbits ($e > 0.5$). 
In contrast, our Atari disk sample extends to higher Galactocentric distances and contains a considerable number of stars with near-circular orbits. Hence, the Kraken appears to be unrelated to the Atari disk.

\textit{Heracles}: Heracles is a stellar population in the inner Milky Way (R$_{GC} < 4$\,kpc) identified from the SDSS/APOGEE survey DR16 \citep{Heracles} with an accretion origin. Looking at the integrals of motion space of Heracles (wide range of orbital energies centered around L$_{z} \approx 0$; see figure~8 in \citealt{Heracles}) and the Atari Disk (narrow range of orbital energies with wide range L$_{z}$), as currently identified, suggests no immediate association. At face value, the less evolved part of Heracles could occupy higher orbital energy values, similar to what we found for the Atari disk. 
However, it is still unclear how to reconcile the discrepant $L_z$ distributions. 
This makes an association between the two populations seem unlikely.

\textit{Nyx}:
Nyx is a prograde (V$_{r} \approx 134$\,km\,s$^{-1}$, V$_{\phi} \approx 130$\,km\,s$^{-1}$, V$_{\theta} \approx 53$\,km\,s$^{-1}$) stellar stream spatially located close to the Galactic disk, originally identified in position-velocity space \citep{Nyx_Lina2020}. 
For a direct comparison with the kinematic properties of Nyx, we calculated the spherical velocity components (V$_{r}$ , V$_{\phi}$, V$_{\theta}$) for our Atari disk sample using Galactocentric spherical coordinates described in Appendix B of \citet{Binney+2008}.
Interestingly, the mean rotational velocity for our Atari disk sample (V$_{\phi} \approx 150$\,km\,s$^{-1}$), somewhat overlaps with the V$_{\phi}$ of Nyx (V$_{\phi} \approx 130$\,km\,s$^{-1}$). 
However, the mean velocity in the radial direction of our Atari disk sample ($\langle\text{V}_{r}\rangle \sim$ 10\,km\,s$^{-1}$) is in stark disagreement with V$_{r} \approx 134$\,km\,s$^{-1}$ for Nyx. 
Nyx also has a mean eccentricity value of $e = 0.68$, somewhat distinct with the eccentricity distribution for the Atari disk. 

Given some similarities in the properties of Nyx and the Atari disk, we further tested the association of Nyx with the Atari disk by compiling the 6-D phase space information of the Nyx sample (Necib, priv. comm.) and ran the sample through our classification algorithm in Section~\ref{sec:selection}.
Essentially all of the Nyx stars are classified as being associated with the Galactic halo, not the Atari disk. 
We thus conclude that the Nyx stellar stream is most likely not associated with the Atari disk. 

We also separately investigated the V$_{r}$ distribution of our Atari disk sample. If this distribution shows two peaks, both with the same mean value but opposite sign, it is a sign of a radial merger event that led to the formation of this structure. At the same time, the two $V_r$ peaks should display near-identical mean rotational velocity and mean azimuthal velocity values. This has been shown to be the case for e.g. the GSE \citep{GSE_Belokurov2018}, Nyx \citep{Nyx_Lina2020} and others. 

To test for this scenario, for a well-defined sample of Atari disk stars within the solar radius ($7 <$\,kpc R $< 9$\,kpc), we performed a 3D Gaussian mixture model fitting over the velocity space (V$_{r}$, V$_{\phi}$, and V$_{\theta}$) of our Atari disk sample (2874 stars). This yielded two Gaussian distributions that peak at V$_{r}$ = 42\,km\,s$^{-1}$ and $-43$\,km\,s$^{-1}$. These two peaks also have V$_{\phi}$= 145\,km\,s$^{-1}$ and 150\,km\,s$^{-1}$, V$_{\theta}$= 0.67\,km\,s$^{-1}$ and 0.95\,km\,s$^{-1}$, respectively.
The surprisingly good match of the two peaks' values of V$_{r}$, V$_{\phi}$, and  V$_{\theta}$ strongly suggest that the Atari disk was formed through a radial merger.

Finally, we compared the location of our Atari disk sample in $E-L_{z}$ to the location of other galactic structures, following the right panel of figure~4 in \citet{Naidu2020}. 
For this comparison, we recalculated the orbital energy of our Atari disk sample and Nyx stream, as described in \citet{Naidu2020} for consistency. 
Figure~\ref{fig:E_Lz_space} shows a schematic view of the approximate location in $E-L_{z}$ space of our Atari disk sample (blue ellipsoid) and the other stellar structures. 
We also added the approximate location of Nyx (dark cyan ellipsoid). 

Notably, the Atari disk does not overlap with the Kraken, the Helmi stream, and the GSE. 
However, it partially overlaps with Wukong, the in-situ halo, and the rich-$\alpha$-disk (aka. Thick Disk). 
But the wider spread in the orbital energy and L$_z$ of our Atari disk sample rules out any close association with the in-situ halo or the Wukong structures. 
Also, the rotational velocity lag of our Atari disk sample with respect to the TD rules out a full association with the component. 
Overall, the unique location of the Atari disk in $E-L_{z}$ space further supports its definition as its own component or structure within the Milky Way.

We conclude that the Atari disk is a unique, ancient low-metallicity component that is located within the Galactic disk that is not associated with any other currently known structure due to its distinct properties (e.g., velocities, eccentricities, metallicities, and location in $E-L_{z}$). 
Looking ahead, it will be important to further study this component.
More detailed information on the chemical abundances of Atari disk stars, especially those with $\metal<-3.0$ could reveal meaningful insights into the nature of the progenitor system (mass, star formation history, accretion time) that may have formed quickly and grown significantly over a short period before merging by the proto-Milky Way. 
All massive structures identified in the Galactic halo, such as the GSE, seems to account for the total mass from which the early Milky Way grew. However, no mass estimates were considering any additional structures potentially hiding in the disk \citep{Naidu2020,Kraken2020}. It thus appears that the Atari disk adds to the observed tally of galactic structures with massive progenitors that will need to be taken into consideration when establishing the early accretion history of the Milky Way.\newline

\section{Summary} 
\label{sec:conclusion}

In this extensive chemo-dynamic study, we have comprehensively characterized the Atari disk as a separate component of the Milky Way's disk. 
Below, we highlight our main conclusions regarding the nature and origin of the Atari disk:

\begin{itemize}
\item We developed a dynamical approach to statistically assign 36,010 low-metallicity stars selected from SkyMapper DR2 to the Galactic thin disk, thick disk and halo populations. 
We utilized two independent probability distribution function approaches using the action integrals and a velocity-based method (following \citet{Bensby2003}) to isolate a clean Atari disk sample while also minimizing the contamination by Galactic halo members, and thin and thick disk stars. 
Our clean Atari disk sample comprises 7,127 stars, all with $-3.5<$ \metal $< -0.8$.

\item We find the Atari disk to have a scale length of $h_{R} = 2.48\,\pm\,0.15$\,kpc and scale height of $h_{Z} = 1.67\,\pm\,0.15$\,kpc. The metallicity distribution of the Atari disk has notable correlations with $|Z|$, $V_{\phi}$, $e$, and $R$. The Atari disk sample shows a mean rotational Velocity of V$_{\phi} \approx 154$\,km\,s$^{-1}$ and a broad eccentricity distribution that peaks at $e = 0.45$. 
The Atari disk sample has a number of stars with higher eccentricity orbits than the canonical thick disk. 
It remains to be seen to what extent the scale length and scale height are dependent on metallicity. 

\item Based on our understanding of the nature of the Atari disk and the properties of our sample, we also developed a simple recipe that could be readily applied to any sample to single out Atari disk stars.

\item  Utilizing photometric metallicities adopted from \citet{Chiti2021} (in combination with high quality \textit{Gaia} EDR3 astrometric solutions), in our clean Atari disk sample of of 7,127 stars, we identify 261 stars with {\metal} $<- 2.0$, 66 stars with $\metal \lesssim -2.5$, and 11 stars with $\metal \lesssim -3.0$. 
Also, through an additional search, we find 17 stars with $\metal \lesssim -3.0$ and five stars with $\metal \lesssim -4.0$ in the literature (collected through JINAbase) to be associated with the Atari disk. 
All these metallicities are below the long-standing metallicity floor of $ {\metal} = -2.35$ (\citealt{Beers2002}) of the thick disk. 
In fact, the discovery of these extremely and ultra-metal-poor stars opens a window to studying the nature and formation history of the proto-disk of our Galaxy.

\item Comparing our results with predictions from the four popular formation scenarios for the formation and evolution of the thick disk (disk heating, gas-rich merger, direct accretion, and radial migration), we conclude that the Atari disk may have been formed through accretion, analogous to what has been suggested for the canonical thick disk direct accretion scenario. Significant roles played by other mechanisms in forming the Atari disk are observationally disfavored. 
This strongly argues for the need for tailored models to attempt to explain the observed properties to further reveal the origin and history of the Atari disk, and its relation to the other disk components.

\item We quantified the shape of the MDF for our Atari disk sample. 
It is well fit by exponential profile with a slope of $\Delta\log {\rm N} / \Delta \metal = 1.13 \pm 0.06$ over the entire metallicity range of our sample, reaching down to $\metal \sim -4.0$, in line with several ultra-metal-poor stars being identified as members of the Atari disk. 
The MDF currently shows no clear peak, which may be caused by the likely increasing member contamination by canonical thick disk stars as [Fe/H] becomes higher than $\sim-1.0$.
The mass of the Atari disk is likely lower than $\sim10^9$\,M$_\odot$, both from the fact that it ought to be dwarfed by the canonical thick disk, and from the mass-metallicity relation assuming an upper bound of $\langle$[Fe/H]$\rangle$ = $-0.8$.

\item We have investigated the existence of any direct association of our Atari disk component with the following Milky Way structures: \textit{Gaia-Sausage-Enceladus}, \textit{Kraken}, and \textit{Nyx}, through comparing their space parameters and properties in the $E-L_z$ plane. 
These comparisons suggest no strong evidence that the Atari disk is associated with other Galactic structures.

\item This study opens a window for the need of more extensive formation modeling(s) of the Galactic disk system and its history, cosmological simulations of the early Milky Way, and precise future observations of Atari disk stars. All these approaches will be required to further investigate in even more detail the observed chemo-dynamical properties of the Atari disk to comprehensively reconstruct its origin scenario and subsequent evolution. Quantifying its role within the early formation of the Galaxy will have important ramifications in understanding the history of our Milky Way.

\end{itemize}

\acknowledgements

We thank John E. Norris, Alexander Ji, Lina Necib, Tilman Hartwig, Miho Ishigaki, Chengdong Li, and Oudai Oweis for fruitful discussions about stellar populations.
This work is supported by Basic Research Grant (Super AI) of Institute for AI and Beyond of the University of Tokyo. A.F. acknowledges support from NSF grant AST-1716251, and thanks the Wissenschaftskolleg zu Berlin for their wonderful Fellow's program and generous hospitality. 

This work has made use of data from the European Space Agency (ESA) mission
{\it Gaia} (\url{https://www.cosmos.esa.int/gaia}), processed by the {\it Gaia}
Data Processing and Analysis Consortium (DPAC,
\url{https://www.cosmos.esa.int/web/gaia/dpac/consortium}). Funding for the DPAC
has been provided by national institutions, in particular the institutions
participating in the {\it Gaia} Multilateral Agreement.

The national facility capability for SkyMapper has been funded through ARC LIEF grant LE130100104 from the Australian Research Council, awarded to the University of Sydney, the Australian National University, Swinburne University of Technology, the University of Queensland, the University of Western Australia, the University of Melbourne, Curtin University of Technology, Monash University and the Australian Astronomical Observatory. SkyMapper is owned and operated by The Australian National University's Research School of Astronomy and Astrophysics. The survey data were processed and provided by the SkyMapper Team at ANU. The SkyMapper node of the All-Sky Virtual Observatory (ASVO) is hosted at the National Computational Infrastructure (NCI). Development and support the SkyMapper node of the ASVO has been funded in part by Astronomy Australia Limited (AAL) and the Australian Government through the Commonwealth's Education Investment Fund (EIF) and National Collaborative Research Infrastructure Strategy (NCRIS), particularly the National eResearch Collaboration Tools and Resources (NeCTAR) and the Australian National Data Service Projects (ANDS).

Funding for RAVE has been provided by the Australian Astronomical Observatory; the Leibniz-Institut fuer Astrophysik Potsdam (AIP); the Australian National University; the Australian Research Council; the French National Research Agency; the German Research Foundation (SPP 1177 and SFB 881); the European Research Council (ERC-StG 240271 Galactica); the Istituto Nazionale di Astrofisica at Padova; The Johns Hopkins University; the National Science Foundation of the USA (AST-0908326); the W. M. Keck foundation; the Macquarie University; the Netherlands Research School for Astronomy; the Natural Sciences and Engineering Research Council of Canada; the Slovenian Research Agency; the Swiss National Science Foundation; the Science $\&$ Technology Facilities Council of the UK; Opticon; Strasbourg Observatory; and the Universities of Groningen, Heidelberg and Sydney.

This work made use of the Third Data Release of the GALAH Survey \citep{Buder+2021}. The GALAH Survey is based on data acquired through the Australian Astronomical Observatory, under programs: A/2013B/13 (The GALAH pilot survey); A/2014A/25, A/2015A/19, A2017A/18 (The GALAH survey phase 1); A2018A/18 (Open clusters with HER- MES); A2019A/1 (Hierarchical star formation in Ori OB1); A2019A/15 (The GALAH survey phase 2); A/2015B/19, A/2016A/22, A/2016B/10, A/2017B/16, A/2018B/15 (The HERMES-TESS program); and A/2015A/3, A/2015B/1, A/2015B/19, A/2016A/22, A/2016B/12, A/2017A/14 (The HERMES K2-follow-up program). We acknowledge the traditional owners of the land on which the AAT stands, the Gamilaraay people, and pay our respects to elders past and present. This paper includes data that has been provided by AAO Data Central (datacentral.aao.gov.au).

The Guoshoujing Telescope (the Large Sky Area Multi-Object Fiber Spectroscopic Telescope; LAMOST) is a National Major Scientific Project built by the Chinese Academy of Sciences. Funding for the project has been provided by the National Development and Reform Commission. LAMOST is operated and managed by the National Astronomical Observatories, Chinese Academy of Sciences.

This research has made use of the SIMBAD database,
operated at CDS, Strasbourg, France. 

\bibliography{Disks}
\end{document}